\documentclass[twocolumn,iop,numberedappendix,appendixfloats]{openjournal}
\usepackage[utf8]{inputenc}
\usepackage{amsmath,amssymb,amstext}
\usepackage{graphicx}
\graphicspath{ {./images/} }
\usepackage{natbib}
\usepackage{subcaption}
\usepackage[colorlinks,allcolors=blue]{hyperref}
\usepackage{dsfont}
\usepackage{xcolor}
\usepackage{listings}
\usepackage{fontawesome}
\definecolor{maroon}{cmyk}{0, 0.87, 0.68, 0.32}
\definecolor{halfgray}{gray}{0.55}
\definecolor{ipython_frame}{RGB}{207, 207, 207}
\definecolor{ipython_bg}{RGB}{247, 247, 247}
\definecolor{ipython_red}{RGB}{186, 33, 33}
\definecolor{ipython_green}{RGB}{0, 128, 0}
\definecolor{ipython_cyan}{RGB}{64, 128, 128}
\definecolor{ipython_purple}{RGB}{170, 34, 255}

\lstdefinelanguage{iPython}{
    morekeywords={access,and,break,class,continue,def,del,elif,else,except,exec,finally,for,from,global,if,import,in,is,lambda,not,or,pass,print,raise,return,try,while},%
    morekeywords=[2]{abs,all,any,basestring,bin,bool,bytearray,callable,chr,classmethod,cmp,compile,complex,delattr,dict,dir,divmod,enumerate,eval,execfile,file,filter,float,format,frozenset,getattr,globals,hasattr,hash,help,hex,id,input,int,isinstance,issubclass,iter,len,list,locals,long,map,max,memoryview,min,next,object,oct,open,ord,pow,property,range,raw_input,reduce,reload,repr,reversed,round,set,setattr,slice,sorted,staticmethod,str,sum,super,tuple,type,unichr,unicode,vars,xrange,zip,apply,buffer,coerce,intern, function, @model, Uniform, Normal, MvNormal, theory_planck},%
    sensitive=true,%
    morecomment=[l]\#,%
    morestring=[b]',%
    morestring=[b]",%
    morestring=[s]{'''}{'''},
    morestring=[s]{"""}{"""},
    morestring=[s]{r'}{'},
    morestring=[s]{r"}{"},%
    morestring=[s]{r'''}{'''},%
    morestring=[s]{r"""}{"""},%
    morestring=[s]{u'}{'},
    morestring=[s]{u"}{"},%
    morestring=[s]{u'''}{'''},%
    morestring=[s]{u"""}{"""},%
    literate=
    {á}{{\'a}}1 {é}{{\'e}}1 {í}{{\'i}}1 {ó}{{\'o}}1 {ú}{{\'u}}1
    {Á}{{\'A}}1 {É}{{\'E}}1 {Í}{{\'I}}1 {Ó}{{\'O}}1 {Ú}{{\'U}}1
    {à}{{\`a}}1 {è}{{\`e}}1 {ì}{{\`i}}1 {ò}{{\`o}}1 {ù}{{\`u}}1
    {À}{{\`A}}1 {È}{{\'E}}1 {Ì}{{\`I}}1 {Ò}{{\`O}}1 {Ù}{{\`U}}1
    {ä}{{\"a}}1 {ë}{{\"e}}1 {ï}{{\"i}}1 {ö}{{\"o}}1 {ü}{{\"u}}1
    {Ä}{{\"A}}1 {Ë}{{\"E}}1 {Ï}{{\"I}}1 {Ö}{{\"O}}1 {Ü}{{\"U}}1
    {â}{{\^a}}1 {ê}{{\^e}}1 {î}{{\^i}}1 {ô}{{\^o}}1 {û}{{\^u}}1
    {Â}{{\^A}}1 {Ê}{{\^E}}1 {Î}{{\^I}}1 {Ô}{{\^O}}1 {Û}{{\^U}}1
    {œ}{{\oe}}1 {Œ}{{\OE}}1 {æ}{{\ae}}1 {Æ}{{\AE}}1 {ß}{{\ss}}1
    {ç}{{\c c}}1 {Ç}{{\c C}}1 {ø}{{\o}}1 {å}{{\r a}}1 {Å}{{\r A}}1
    {€}{{\EUR}}1 {£}{{\pounds}}1
    {^}{{{\color{ipython_purple}\^{}}}}1
    {=}{{{\color{ipython_purple}=}}}1
    {+}{{{\color{ipython_purple}+}}}1
    {-}{{{\color{ipython_purple}-}}}1
    {*}{{{\color{ipython_purple}$^\ast$}}}1
    {/}{{{\color{ipython_purple}/}}}1
    {+=}{{{+=}}}1
    {-=}{{{-=}}}1
    {*=}{{{$^\ast$=}}}1
    {/=}{{{/=}}}1,
    literate=
    *{-}{{{\color{ipython_purple}-}}}1
     {?}{{{\color{ipython_purple}?}}}1,
    identifierstyle=\color{black}\ttfamily,
    commentstyle=\color{ipython_cyan}\ttfamily,
    stringstyle=\color{ipython_red}\ttfamily,
    keepspaces=true,
    showspaces=false,
    showstringspaces=false,
    rulecolor=\color{ipython_frame},
    frameround={t}{t}{t}{t},
    numbers=none,
    numberstyle=\tiny\color{halfgray},
    backgroundcolor=\color{ipython_bg},
    basicstyle=\ttfamily\footnotesize,
    columns=fullflexible,
    keywordstyle=\color{ipython_green}\ttfamily,
}

\newcommand{\capse}{\texttt{Capse.jl}}
\newcommand{\camb}{\texttt{CAMB}}
\newcommand{\cobaya}{\texttt{Cobaya}}
\newcommand{\planck}{\textit{Planck}}
\newcommand{\plancklite}{\texttt{PlanckLite.jl}}
\newcommand{\julia}{\texttt{Julia}}
\newcommand{\turing}{\texttt{Turing.jl}}
\newcommand{\classnet}{\texttt{ClassNet}}
\newcommand{\cosmopower}{\texttt{CosmoPower}}

\newcommand{\github}{\href{https://github.com/marcobonici/capse_paper}{\faGithub}}

\date{\today}

\begin{document}
\journalinfo{The Open Journal of Astrophysics}
\submitted{submitted XXX; accepted YYY}

\shorttitle{\capse{}: Efficient and Auto-Differentiable CMB power spectra emulation}
\shortauthors{Bonici, Bianchini \& Ruiz-Zapatero}
\title{\capse{}: Efficient and Auto-Differentiable CMB power spectra emulation}
\author{Marco Bonici$^{\star1, 2, 3}$}
\author{Federico Bianchini$^{\dagger 4, 5, 6}$}
\author{Jaime Ruiz-Zapatero$^{\ddagger 7}$}
\affiliation{$^1$ INAF-IASF Milano IT}
\affiliation{$^2$ Waterloo Centre for Astrophysics, University of Waterloo, Waterloo, ON N2L 3G1, Canada}
\affiliation{$3$ Department of Physics and Astronomy, University of Waterloo, Waterloo, ON N2L 3G1, Canada}
\affiliation{$^4$ Kavli Institute for Particle Astrophysics and Cosmology, Stanford University, 452 Lomita Mall, Stanford, CA, 94305, USA}
\affiliation{$^5$SLAC National Accelerator Laboratory, 2575 Sand Hill Road, Menlo Park, CA 94025}
\affiliation{$^6$Department of Physics, Stanford University, 382 Via Pueblo Mall, Stanford, CA 94305}
\affiliation{$^7$Astrophysics, University of Oxford, DWB, Keble Road, Oxford OX1 3RH, UK}
\thanks{$^\star$ E-mail: \nolinkurl{marco.bonici@inaf.it} \\
$\dagger$ E-mail: \nolinkurl{fbianc@stanford.edu} \\ $\ddagger$ E-mail: \nolinkurl{jaime.ruiz-zapatero@physics.ox.ac.uk}}

\begin{abstract}
We present \capse{}, a novel neural network-based emulator designed for rapid and accurate prediction of Cosmic Microwave Background (CMB) temperature, polarization, and lensing angular power spectra. The emulator computes predictions in just a few microseconds with emulation errors below $0.1\sigma$ for all the scales relevant for the upcoming CMB-S4 survey. \capse{} can also be trained in an hour's time on a 8-cores  CPU.
We test \capse{} on \planck{} 2018, ACT DR4, and 2018 SPT-3G data and demonstrate its capability to derive cosmological constraints comparable to those obtained by traditional methods, but with a computational efficiency that is three to six orders of magnitude higher. We take advantage of the differentiability of our emulators to use gradient-based methods, such as Pathfinder and Hamiltonian Monte Carlo (HMC), which speed up the convergence and increase sampling efficiency. 
Together, these features make \capse{} a powerful tool for studying the CMB and its implications for cosmology. When using the fastest combination of our likelihoods, emulators, and analysis algorithm, we are able to perform a \planck{} $TT+TE+EE$ analysis in less than a second. To ensure full reproducibility, we provide open access to the codes and data required to reproduce all the results of this work \github
\end{abstract}

\maketitle
\section{Introduction}
\label{sec:introduction}

The experimental cosmological landscape is extremely rich and promising.
Upcoming millimeter-wave telescopes, including the Simons Observatory \citep[SO,][]{Ade_2019} and CMB-S4 \citep{abazajian2019cmbs4}, are building on the legacy of the \planck{} satellite \citep{Planck:2019nip}, the South Pole Telescope \citep[SPT,][]{Sobrin_2022} and the Atacama Cosmology Telescope \citep[ACT,][]{aiola_atacama_2020} to deliver precise, sample variance dominated measurements of cosmic microwave background (CMB) anisotropies down to arcminute scales in intensity and polarization. 
Meanwhile, photometric and spectroscopic galaxy surveys such as the Dark Energy Survey~\citep{abbott_dark_2018}, KiDS~\citep{de_jong_kilo-degree_2013}, BOSS~\citep{dawson_baryon_2013}, DESI~\citep{levi_desi_2013}, Euclid~\citep{laureijs_euclid_2011}, the Vera Rubin Observatory~\citep{lsst_dark_energy_science_collaboration_large_2012}, the Nancy Grace Roman Space Telescope~\citep{spergel_wide-field_2015}, and SPHEREx~\citep{crill_spherex_2020} are poised to accurately measure the positions and shapes of billions of galaxies, reconstructing the large-scale structure (LSS) of the Universe out to high redshifts. 
Fueled by a dramatic increase in the data throughput, these experiments promise to map out a large portion of the Hubble volume to put under stress the $\Lambda$CDM model, delivering groundbreaking insights into the fundamental properties of the cosmos.

However, analyzing the large amounts of data generated by these experiments poses significant challenges. 
In most cases, the analysis is carried out in terms of summary statistics, such as the two-point function in either real or harmonic space. 
At the heart of Bayesian inference pipelines, Einstein-Boltzmann codes such as \camb{}~\citep{lewis_efficient_2000} or \texttt{CLASS}~\citep{blas_cosmic_2011} solve for the evolution of cosmological perturbations and calculate CMB/LSS observables, which are in turn compared to the data in the likelihood evaluation.
Depending on the dimensionality and the topology of the parameter space, this operation is typically repeated $10^4$-$10^6$ times before convergence thresholds are reached; considering that a call to a Boltzmann solver might require from a second to a minute in case of high-precision settings, these numbers show the high computational cost of Monte Carlo Markov Chains (MCMC) analysis.

Emulation has emerged as a highly promising technique for accelerating cosmological inference.
It consists in developing surrogate models that approximate the outputs of computationally expensive models at significantly lower computational costs. This approach offers a twofold advantage: it expedites calculations and enables the utilization of alternative optimization and sampling methods, such as those based on gradients. 
In recent years, emulators have proven successful in various domains with computationally intensive forward models, including cosmology.
Early contributions in this area can be traced back to the work of \cite{jimenez_fast_2004} and \cite{fendt_pico_2007},  which presented the initial applications of emulators based on polynomial interpolation within the field of cosmology.  
The first emulator based on neural networks (NN), called \texttt{CosmoNet}, was introduced in \cite{auld_fast_2007}. 
Recent advancements include the work of \texttt{CosmicNet}~\citep{albers_cosmicnet_2019, gunther_cosmicnet_2022}, with emulators targeting the replacement of the most computationally demanding components within Boltzmann solvers. 
Additionally, \cite{nygaard_connect_2022}, \cite{Gunther:2023xhh}, and \cite{To:2022ubu} adopted an iterative approach for emulator training to reduce the overall computational resources required.
Currently, the most precise CMB emulator (\textit{i.e.} matching the \camb{} high-precision settings) that accurately coveres a wide parameter space is \cosmopower{}~\citep{mancini_itcosmopower_2021, bolliet_high-accuracy_2023}, which directly emulates the CMB angular power spectra.

The cosmological community also developed emulators tailored for LSS analysis, with emulators for the linear matter power spectrum~\citep{Mootoovaloo:2021rot, mancini_itcosmopower_2021, donald-mccann_textttmatryoshka_2021, arico_accelerating_2021}, nonlinear corrections~\citep{Lawrence_2017, DeRose_2019, euclid_collaboration_euclid_2019, euclidcollaboration_euclid_2021, angulo_bacco_2021}, power spectrum multipoles~\citep{arico_accelerating_2021, eggemeier_comet_2022, donald-mccann_textttmatryoshka_2022}, galaxy survey angular power spectrum~\citep{manrique-yus_euclid-era_2019, mootoovaloo_parameter_2020, bonici_fast_2022, To:2022ubu} and even for the log-posterior~\citep{Gammal:2022eob}. The amount of papers employing emulators is rapidly growing, demonstrating an increasing interest of the cosmological community toward this technique.

In this paper, we present a novel emulator of the CMB temperature, polarization, and lensing potential anisotropies angular power spectra, \capse{}\footnote{\href{https://github.com/CosmologicalEmulators/Capse.jl}{\texttt{https://github.com/CosmologicalEmulators/Capse.jl}}}, which differs from previously developed emulators in two fundamental ways. On the one hand, we propose and implement a more efficient preprocessing, \textit{i.e.} the operations performed on the training data before being fed to the NN for the training, and  test a different technique for data compression. On the other hand, we employ a specialized, performant NN framework in \julia{}, which results in our emulator returning predictions in around 50 $\mu s$.
Moreover, \capse{} can be trained in an hour's time on a standard 8-cores CPU to cover a wide range of cosmological parameters, and down to scales that will be mapped by upcoming experiments. These characteristics make \capse{} the CMB power spectra emulator with the shortest evaluation time available as well as extremely cheap to train. In addition, we also implement fully differentiable CMB likelihoods. Coupled to our emulators, these methods enable the use of gradient-based inference methods. The combination of these state-of-the-art inference methods with \capse{} allows us to analyze \planck{} data in less than one second and ACT data in $\sim10$ seconds.

This paper is structured as follows. In Sec.~\ref{sec:capse_description} we describe the architecture, the preprocessing, and the training procedure of our emulators.
In Sec.~\ref{sec:likelihood} we describe the way we construct our likelihoods. In Sec.~\ref{sec:samplers} we describe the sampling algorithms employed in this work. In Sec.~\ref{sec:results} we show the main results of our work, namely the precision and the speed of our emulators coupled with differentiable likelihoods. In Sec.~\ref{sec:comparison} we compare our emulators with similar works present in the literature. Finally, we conclude drawing the conclusions of this study in Sec.~\ref{sec:conclusions}.

\section{Capse emulator}
\label{sec:capse}
The core idea behind surrogate models is quite simple: replacing a computationally expensive function with a sufficiently accurate approximation that is cheaper. 
Although creating an emulator is conceptually straightforward, there are several design choices involved in the  process which can have a drastic impact on the final performance. 
These choices include the type and architecture of the surrogate model, the preprocessing procedure, the application of dimensional reduction techniques, and potentially incorporating domain knowledge.

To achieve optimal performance and train a fast and accurate emulator while conserving computational resources, all of these choices become significant. In this section, we provide a detailed description of our \capse{} emulator, focusing on the aspects that differentiate it from other emulators present in literature. We conclude by discussing the accuracy of the trained emulators using a validation data  set that covers the entire range of emulation.

\label{sec:capse_description}
\subsection{Architecture, preprocessing and training}
\label{sec:architecture}
The NN employed for our emulators is a multi-layer perceptron (MLP) with a number of input features that corresponds to the number of cosmological parameters considered, 5 hidden layers with 64 neurons each and a hyperbolic tangent activation function.
Our emulators are written in the \julia{} language~\citep{bezanson_julia_2015} and employ the NN implemented in \texttt{SimpleChains.jl}.
While frameworks like \texttt{PyTorch}, \texttt{TensorFlow}, and \texttt{JAX} offer general-purpose capabilities to cover a wide range of use cases, from simple MLPs to complex Large Language Models and from models running on a single CPU to those using grids of TPUs, \texttt{SimpleChains.jl} has a way more limited focus: it excels at running small NNs on CPUs.
By narrowing down the scope, the library prioritizes performance over generality, enabling several choices that improve computational efficiency.
This results in a significant boost compared to more traditional frameworks, as demonstrated in previous comparisons with \texttt{SimpleChains.jl}.\footnote{A more detailed description for the interested reader can be found \href{https://julialang.org/blog/2022/04/simple-chains/}{here}.} 

A key aspect of our methodology is the preprocessing of training data, which significantly impacts emulation performance. In this work, we perform min-max normalization on each input and output feature. This process scales all features to fall within a $[0,1]$ range, a technique known to enhance emulation performance ~\citep{nygaard_connect_2022}. 
Furthermore, we leverage domain knowledge, exploiting the dominant linear dependence of the $C_\ell$'s on the scalar amplitude $A_s$. 
To account for this, we divide each output feature for the corresponding value of $A_s$, while still treating $A_s$ as an independent parameter within the NN.
This approach analytically handles the linear scaling, allowing the NN to focus on learning the more  nonlinear feature.\footnote{A similar rescaling for $\tau$ could potentially be employed. However, as its effect varies across different power spectra and scales~\citep{Alvarez:2020gvl} and since the desired accuracy was achieved with our current procedure, further optimization for $\tau$ is left for future work.} 
We found that this enhances the emulation accuracy given the same amount of training resources. 
This improvement can be easily explained by the fact that, given the $\ln 10^{10}A_s$ emulation range, these preprocessing steps reduce the range of the output features by almost a factor of 3, hence improving the emulation performance. We checked it explicitly: the $A_s$ rescaling enhances the emulator accuracy by a factor of $\approx 3$, almost the same as the $C_\ell$'s dynamical range reduction.

The emulation space is sampled with a Latin-Hypercube Sampling (LHS) algorithm~\citep{McKay1979} using 10,000 points. 
The ground-truth predictions have been computed using the \camb{} Boltzmann solver~\citep{lewis_efficient_2000} with the same high-accuracy settings of~\cite{bolliet_high-accuracy_2023}. In particular, we set \texttt{lens\_potential\_accuracy}=8, \texttt{lens\_margin}=2050, \texttt{AccuracyBoost}=2.0, \texttt{lSampleBoost}=2.0, \texttt{lAccuracyBoost}=2.0 and \texttt{DoLateRadTruncation}=\texttt{False}. The nonlinear corrections were computed using \texttt{HMCode}~\citep{mead_hmcode-2020_2021}.
The emulation space is also chosen to exactly match that of \cite{bolliet_high-accuracy_2023}, which allows us to make a fair comparison with their results; the details can be found in Table~\ref{tab:emulation_space}.
We trained a NN for each angular spectrum considered in this work (namely $TT,\,TE,\,EE,\,\phi\phi$), using a standard laptop with an 8-core CPU using the \texttt{ADAM} optimizer~\citep{kingma_adam_2017} with a mean square error loss function for 50,000 epochs.

In this paper, we explore two different emulation approaches. In the first approach, the output of the NN is a set of CMB power spectra $C_\ell$'s evaluated on the multipole grid from $\ell_\mathrm{min}=2$ to $\ell_\mathrm{max}=5000$ with a step of $\Delta\ell = 1$. 
This emulator delivers predictions for a single power spectrum in a meantime of $40\,\mu$s.
This improvement is truly remarkable when compared to the computation time of predictions using a Boltzmann solver with high-accuracy settings, which typically takes around $\mathcal{O}(60\,\text{s})$. Consequently, our emulators achieve speeds that are approximately 1,000,000 times faster.

The second approach we consider involves a dimensionality reduction technique, the Chebyshev polynomial decomposition. The key idea behind this approach is to project the $\ell$-dependence of the $C_\ell$'s onto the Chebyshev polynomial basis:
\begin{equation}
    C_\ell(\theta)\approx\sum_{n=0}^{N_\mathrm{max}} a_n(\theta)T_n(\ell),
    \label{eq:chebyshev}
\end{equation}
where $a_n$ represents the $n$-th coefficient that carries the dependence on the cosmological parameters $\theta$, $T_n$ denotes the $n$-th Chebyshev polynomial, and $N_\mathrm{max}$ is the highest degree of the polynomials employed in the decomposition.\footnote{Note that we use a rescaled version of the Chebyshev polynomials since they are defined on the interval $\left[-1, 1\right]$. The rescaling is automatically handled by the Chebyshev projection library employed in our analysis, \href{https://github.com/JuliaMath/FastChebInterp.jl}{\texttt{FastChebInterp.jl}}}
The emulator is then trained to reproduce the cosmological dependence of the $a_n(\theta)$ coefficients. We choose to explore this approach for the nice mathematical properties of Chebyshev interpolation; we refer the interested reader to~\cite{atap} and~\cite{wang2023analysis} for a detailed discussion.

The advantages of this polynomial approach are twofold. 
First, it significantly speeds up the training process, since the number of output features of the NN is reduced.
We have found that using a polynomial order of $N_\mathrm{max}=48$ provides accurate emulation while simplifying the complexity of the neural network.
This reduction in complexity not only lowers the training cost but also improves the emulator speed, enabling predictions to be generated in approximately $15\,\mu$s. 
This represents a 2.5-fold increase in speed compared to the previous scenario.

The second, perhaps less obvious, advantage stems from this simple observation. 
By performing a linear decomposition of the $C_\ell$'s and noting that the basis functions do not depend on the cosmological parameters, we can leverage this linear decomposition in the likelihood computation. 
If the operations needed to perform the likelihood computation can be represented as a linear operator $\mathds{L}$ acting on the theory vector, we can utilize the linear decomposition as follows:
\begin{equation}
    \mathds{L}\sum_n a_n(\theta) T_n=\sum_n a_n(\theta)\mathds{L}T_n\equiv\sum_na_n\hat{T}_n,
    \label{eq:algebra_trick}
\end{equation}
where we have defined:
\begin{equation}
    \hat{T}_n\equiv\mathds{L}T_n.
\end{equation}
If the linear operator $\mathds{L}$ does not depend on the cosmological parameters, as it typically is the case, we can compute the action of $\mathds{L}$ on the basis function once and store the results. 
As we will demonstrate later, this approach leads to significant speed improvements in the analysis.
\begin{table}
      \centering
      \def\arraystretch{1}
\begin{tabular}{|c|cc|}
\hline Parameter & Min & Max \\
\hline $\ln 10 A_{\mathrm{s}}$ & 2.5 & 3.5 \\
$n_{\mathrm{s}}$ & 0.88 & 1.06 \\
$H_0[\mathrm{km}/s/\mathrm{Mpc}]$ & 40 & 100 \\
$\omega_{\mathrm{b}}$ & 0.01933 & 0.02533\\
$\omega_{\mathrm{c}}$ & 0.08 & 0.20\\
$\tau$ & 0.02 & 0.12\\
\hline
\end{tabular}
\caption{The parameter ranges used to generate the Latin hypercube of cosmological parameters for computing the training data. It is important to note that our emulators should only be used within these specified ranges.}
\label{tab:emulation_space}
\end{table}

\subsection{Accuracy tests}
\label{sec:accuracy_tests}
We carry out two types of tests to assess the performance of the emulators. 
The first check involves comparing the emulator's predictions with the corresponding ground truth values. 
This comparison allows us to measure the discrepancy between the emulator and the Boltzmann solver output. In this section, we will thoroughly discuss this validation approach.
The second check involves running a Markov Chain Monte Carlo (MCMC) analysis using specific datasets.
The MCMC chains produced by the emulator are then compared with the chains obtained from the Boltzmann code.
This comparison serves as an additional validation step to assess the emulator's accuracy and reliability. 
The details and results of this MCMC analysis will be presented in Sec.~\ref{sec:results}.

We use \camb{} to generate a validation dataset consisting of 20,000 combinations of the cosmological parameters drawn from the same emulation range of the training dataset (see Tab.~\ref{tab:emulation_space}).
We emphasize that these specific combinations are not present in the training dataset. 
Our objective is to assess the accuracy of the emulators by comparing their predictions to the ground truth values. 
To quantify the emulation error, we calculate the difference between the emulated $C_\ell^{XY}$ and the true $C_\ell^{XY}$, normalized by the corresponding spectrum variance $\sigma_\ell^{XY}$
\begin{equation}
    \frac{\left|C_{\ell, \text { emulated }}^{XY}-C_{\ell, \text { true }}^{XY}\right|}{\sigma_{\ell}^{XY}},
    \label{eq:err_comparison}
\end{equation}
where $XY \in [TT,\,TE,\,EE,\,\phi\phi]$ represent the CMB temperature, polarization, and lensing potential anisotropies power spectra.

\begin{figure*}%
\centering%
\includegraphics[width=.85\textwidth]{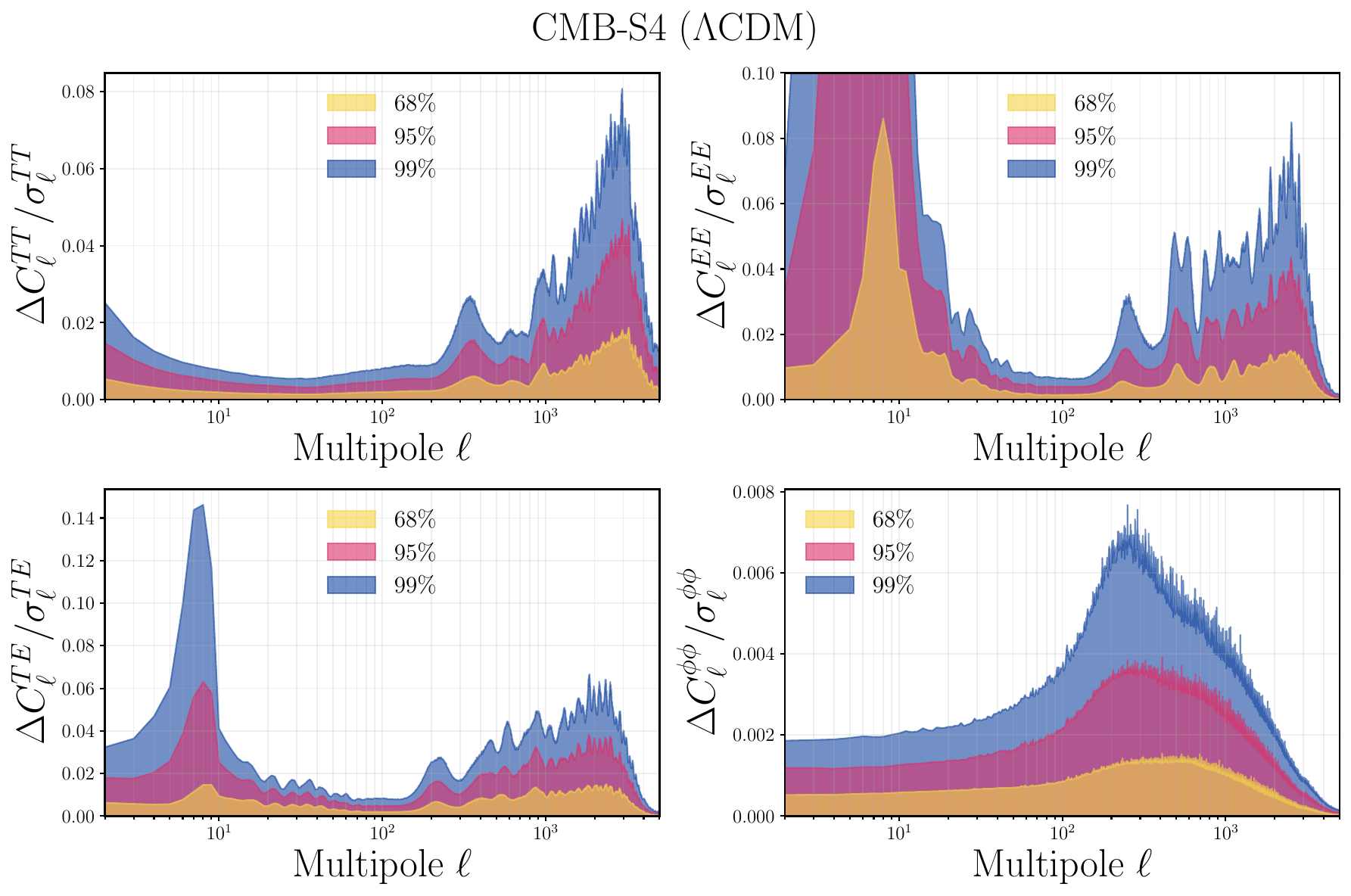}%
\caption{Comparison of the relative differences between \capse{} and high-accuracy predictions from CAMB for the four 2-point statistics emulated in this study. The differences are normalized to the forecasted CMB-S4 statistical uncertainties ($f_\mathrm{sky} = 0.4$) and measured in units of cosmic variance for $\ell < 40$. The shaded regions represent the 68\%, 95\%, and 99\% percentiles of the respective distributions.}%
\label{fig:S4_errors}
\end{figure*}

To perform this test, we need both the ground truth spectra for computing the emulation error and the expected spectra variance $\sigma_\ell$.
The variance can be computed, under the assumption that both $\left[X,\,Y\right]\in\left[{T,\,E,\,\phi}\right]$ behave as Gaussian random fields, as \citep[e.g.,][]{Knox:1995dq,Kamionkowski:1996ks}:
\begin{equation}
    \begin{aligned}
\sigma_{\ell, \mathrm{CMB}}^{XY}= & \sqrt{\frac{1}{f_{\mathrm{sky}}(2 \ell+1)}} \\
& \times \sqrt{(C_{\ell}^{XY} )^2+\left(C_{\ell}^{XX}+N_{\ell}^{XX}\right)\left(C_{\ell}^{YY}+N_{\ell}^{YY}\right)},
\end{aligned}
\label{eq:crosscorr_sigma}
\end{equation}

where $f_\mathrm{sky}$ and $N_{\ell}^{XX}$ are respectively the observed sky fraction and the noise power spectrum of field $X$ for the experimental configuration under consideration.
We consider two representative upcoming ground-based CMB surveys, Simons Observatory\footnote{Available at \url{https://github.com/simonsobs/so_noise_models}. For temperature and polarization we use the following \texttt{LAT\_comp\_sep\_noise/v3.1.0} curves: \texttt{SO\_LAT\_Nell\_T\_atmv1\_goal\_fsky0p4\_ILC\_CMB.txt} and \texttt{SO\_LAT\_Nell\_P\_baseline\_fsky0p4\_ILC\_CMB\_E.txt}. For lensing we choose the \texttt{nlkk\_v3\_1\_0\_deproj0\_SENS2\_fsky0p4\_\\it\_lT30-3000\_lP30-5000.dat} in the \texttt{LAT\_lensing\_noise/lensing\_v3\_1\_1/} folder.} \citep{Ade_2019} and CMB-S4\footnote{The primary CMB noise curves are available at \url{https://sns.ias.edu/~jch/S4_190604d_2LAT_Tpol_default_noisecurves.tgz}, specifically we use \texttt{S4\_190604d\_2LAT\_T\_default\_noisecurves\_deproj0\_SENS0\_mask\_16000\allowbreak\_ell\_TT\_yy.txt} and \texttt{S4\_190604d\_2LAT\_pol\_default\_noisecurves\_deproj0\allowbreak\_SENS0\_mask\_16000\_ell\_EE\_BB.txt} for $TT$ and $EE$ respectively. For the CMB lensing reconstruction noise we use the \texttt{kappa\_deproj0\_sens0\_16000\_lT30-3000\_lP30-5000.dat} curves from \url{https://github.com/toshiyan/cmblensplus/tree/master/example/data}.} \citep{abazajian2019cmbs4}. 
We refer the reader to \cite{bolliet_high-accuracy_2023} and references therein for details regarding the surveys sensitivities.
In both cases, we use projected noise curves obtained after foreground cleaning and including multi-frequency information from the  \planck{} satellite. 
For both experiments we assume an observed sky fraction of $f_{\rm sky}=0.4$; the results can be seen in Figs.~\ref{fig:S4_errors} and \ref{fig:SO_errors}.

The degraded \textit{EE} 2-point function  performance in the $\ell<20$ region arises because of two different factors. 
First and foremost, we only consider cosmic variance over this multipole range since S4 will not be able to image the largest angular scales at $\ell \lesssim 40$ and therefore they do not provide noise curves on this range. 
The second effect comes from the scaling of the 2-point function with respect to $\tau$, which is proportional to $\tau^2$ for $\ell<30$. 
In fact, the dynamical range of the $C_\ell^{EE}$ in this specific multipole range, across all the cosmologies considered, is 30 times bigger than the one outside. We expect that, when including the $\tau$ rescaling in a future work, we will be able to improve the low-$\ell$ accuracy as well.
If we exclude the $C_\ell$'s with $\ell<13$, the results are even better, with an emulation error that is below $0.08\,\sigma$ for $99\%$ of the points in the validation dataset.

\section{Cosmological inference framework}
\label{sec:likelihood}

After providing a detailed description of the \capse{} emulator, our focus shifts to leveraging its CMB power spectra predictions for cosmological inference.
Specifically, we are interested in drawing samples from the posterior distribution of the model parameters $\boldsymbol{\theta}$ given the data $\boldsymbol{d}$, $P(\boldsymbol{\theta}|\boldsymbol{d})$, after specifying a likelihood function $\mathcal{L}(\boldsymbol{d}|\boldsymbol{\theta})$ and  the priors $\Pi(\boldsymbol{\theta})$. 
To this end, we use the \julia{} Bayesian statistical inference library \turing{}~\citep{ge2018t}. 

\begin{table}
      \centering
      \def\arraystretch{1}
      \begin{tabular}{|cc|cc|}
        \hline
        \multicolumn{2}{|c|}{\textbf{Planck}}               &    \multicolumn{2}{c|}{\textbf{ACT}}  \\
        \hline
        Parameter &  Prior & Parameter &  Prior\\  
        \hline 
        $\ln 10A_{\rm{s}}$  &  $U (2.5, 3.5)$                   &     $\ln 10A_{\rm{s}}$  &  $U (2.8, 3.2)$\\
        $n_{\rm{s}}$       & $U (0.88, 1.06)$                 &     $n_{\rm{s}}$       & $U (0.88, 1.06)$\\
        $H_{\rm{0}}$       &  $U (40, 100)$                     &           $H_{\rm{0}}$       &  $U (40, 100)$\\
        $\omega_{\rm{b}}$  & $U (0.01933, 0.02533)$               &     $\omega_{\rm{b}}$  & $U (0.01933, 0.02533)$ \\
        $\omega_{\rm{c}}$  & $U (0.08, 0.20)$                   &     $\omega_{\rm{c}}$  & $U (0.08, 0.20)$ \\
        $\tau$             & $\mathcal{N} (0.0506, 0.0086)$        &     $\tau$             & $\mathcal{N} (0.065, 0.015)$ \\
        $A_\mathrm{Planck}$  & $\mathcal{N} (1.0, 0.001)$         &     $y_{\rm{p}}^2$  & $U (0.7, 1.3)$ \\
        \hline
      \end{tabular}
      \caption{Priors on the cosmological and calibration parameters varied in the analyses of \planck{} and ACT datasets. $U(a,b)$ denotes a uniform distribution between $[a,b]$, while $\mathcal{N}(\mu,\sigma)$ indicates a Gaussian distribution with mean $\mu$ and variance $\sigma^2$.}\label{tab:priors}
\end{table}

\turing{} is a Probabilistic Programming Language (PPL). Unlike traditional programming languages, which are designed for deterministic computations, PPLs are designed to explicitly incorporate probabilistic constructs. This allows the user to explicitly declare the priors for the input parameters and write their relationship to the observations within a statistical mode.  Moreover, the user can also declare the distribution of the observations. Then, \turing{} uses this information to automatically draw an expression for the likelihood by conditioning the model on the observations.

Below is a \julia{} code snippet demonstrating how a likelihood is implemented using \turing{}:
\begin{lstlisting}[language=iPython]
@model function CMB_planck(data, covariance)
    #define the priors
    ln10As   ~ Uniform(2.5, 3.5)
    ns       ~ Uniform(0.88, 1.06)
    H0       ~ Uniform(40, 100)
    omb      ~ Uniform(0.01985, 0.025)
    omc      ~ Uniform(0.08, 0.20)
    tau      ~ Normal(0.0506, 0.0086)
    Aplanck  ~ Normal(1.0, 0.0025)
    #create a parameters vector
    params = [ln10As, ns, H0, omb, omc, tau]
    
    #perform the theoretical computation
    pred = theory_planck(params) ./ (Aplanck^2)

    #compute the likelihood
    data ~ MvNormal(pred, covariance)

    return nothing
end
\end{lstlisting}
This is slightly different from the actual likelihood being used, which performs a reparametrization as explained in Appendix~\ref{app:reparametrization}.

In this work, we use the distributions listed in Tab. \ref{tab:priors} as priors for the input parameters of the \capse{} emulator. The only exception is $\tau$, for which we employ a Gaussian prior, stemming from the analysis of the low-$\ell$ $EE$ angular spectrum.
Moreover, since both \planck{}, ACT, and SPT-3G measurements are Gaussian-distributed (see Sec.~\ref{sec:datasets}), we use \turing{} to draw a Gaussian likelihood of the shape:
\begin{equation}
    \boldsymbol{d}\sim \mathcal{N}(\boldsymbol{t}(\boldsymbol{\theta}), \boldsymbol{C}) \, ,
     \label{eq:lkl}
\end{equation}
where $\boldsymbol{t}(\boldsymbol{\theta})$ is a theory vector computed as a function of the model parameters $\boldsymbol{\theta}$ and $\boldsymbol{C}$ is the covariance matrix of the data vector $\boldsymbol{d}$.

While the bottleneck is typically the computation of the theory vector $\boldsymbol{t}(\boldsymbol{\theta})$, even for the other emulators described in the literature, this is not true anymore in our case, especially for the data and theory vectors of the \planck{} and ACT analyses. The binning of the $C_\ell$'s and the $\chi^2$ computation are more expensive by a factor of 7 than our emulators computations.

In order to overcome this issue, we follow two strategies. The first is the one depicted in Eq.~\ref{eq:algebra_trick}. As anticipated in Sect. \ref{sec:architecture}, all the operations required to compute the \planck{} and ACT likelihoods can be represented as a linear operator $\mathds{L}$ on $\boldsymbol{t}(\boldsymbol{\theta})$. In this case, we can compute its action on the basis function once $\hat{T}_n=\mathds{L}T_n$ and store it. During the MCMC, we multiply $\hat{T}_n$ with $a_n(\boldsymbol{\theta})$ and then compute the $\chi^2$. This procedure, while speeding up the analysis, does not introduce any theoretical error: the computation of the binned $C_\ell$'s using the two approaches matches up to machine-precision.

A further optimization that we can introduce is using
\texttt{MOPED}, a linear data compression algorithm~\citep{Heavens:1999am, Heavens:2017efz}. 
Given a $n$-dimensional data vector whose theoretical prediction depends on $m$ parameters, \texttt{MOPED} generates $n$ $m$-dimensional data vector, $\boldsymbol{b}_i$. 
Moreover, if the noise of the data does not depend on the parameters of the model, as in the case of \planck{} and ACT measurements, the original and compressed data vectors can be shown to have identical Fisher information matrices with respect the parameters of the model~\citep{Heavens:1999am}. 
The \texttt{MOPED} compression vectors $\boldsymbol{b}_i$ can be computed using:
\begin{equation}
    \boldsymbol{b}_i=\frac{\boldsymbol{C}^{-1} \boldsymbol{t}_{,i}-\sum_{j=1}^{i-1}\left(\boldsymbol{t}_{,i}^T \boldsymbol{b}_j\right) \boldsymbol{b}_j}{\sqrt{F_{ii}-\sum_{j=1}^{i-1}\left(\boldsymbol{t}_{,i}^T \boldsymbol{b}_j\right)^2}},
    \label{eq:moped_compression}
\end{equation}
where $\boldsymbol{t}_{,i}$ is the derivative of the theory vector with respect to the $i$-th parameter and $F_{ii}$ is the $i$-th element on the diagonal of the Fisher matrix $\boldsymbol{F}$:
\begin{equation}
    F_{ij}=-\frac{\partial^2 \mathcal{L}(\theta)}{\partial \theta_i \partial \theta_j}.
\end{equation}
As stated in~\cite{Campagne:2023ter}, it is pretty straightforward to evaluate both the Fisher matrix and the $\boldsymbol{t}$ derivatives using automatic differentiation (AD).

Therefore, we use \texttt{MOPED} to perform a compression of the likelihood shown in Eq.~\ref{eq:lkl}. This allows us to transform a relatively expensive $n$-dimensional Gaussian likelihood into the product of $m$ one dimensional unit-variance Gaussian likelihoods: 

\begin{equation} \label{eq:lkl_moped}
    \mathcal{L}(\boldsymbol{d}|\boldsymbol{\theta}) \approx \Pi^m_i \mathcal{N}(\boldsymbol{t}(\boldsymbol{\theta}) M_i, 1) \, ,
\end{equation}

where $M_i$ is the \texttt{MOPED} compression matrix for the $i$-th parameter of the model. However, unlike our first strategy, using the \texttt{MOPED} compression changes the actual likelihood being targeted, opening the possibility for changes the results of our analysis. 

\section{Samplers}
\label{sec:samplers}
We now outline the methods employed to explore the likelihoods introduced in Sec.~\ref{sec:likelihood} and derive parameter constraints for the \planck{}, ACT, and SPT-3G datasets. 
The algorithms employed, all based on leveraging the likelihood gradient, are:
\begin{itemize}
    \item NUTS, a version of the Hamiltonian Monte Carlo sampler~\citep{hoffman2011nouturn}. This is considered the state-of-the-art algorithm to be employed in MCMC analysis.
    \item MicroCanonical Hamiltonian Monte Carlo, a recently proposed extension of Hamiltonian MonteCarlo that promises to be competitive with NUTS~\citep{robnik2022microcanonical, robnik2023microcanonical}.
    \item Pathfinder, a variational inference methods~\citep{zhang2022pathfinder}. Although less robust than the previous methods, as it lacks of asymptotic convergence properties, Pathfinder excels at quickly delivering draws from the typical set, that can be representative of the posteriors in the case of Gaussian or mildly non-Gaussian posterior, or used to initialize standard MCMC.
\end{itemize}

In this section we describe all these methods, while addressing the interested reader to the relevant works in the literature.

\subsection{Hamiltonian MonteCarlo}
\label{sec:hmc}
Hamiltonian Monte Carlo (HMC)~\citep{betancourt2018conceptual} is a MCMC algorithm that explores a parameter space by simulating the dynamics of a Hamiltonian system. The fundamental idea behind HMC is to introduce auxiliary momentum variables that are independent of the target function. The joint distribution of the position (\textit{i.e.} the original parameters) and momentum variables is then defined by a Hamiltonian function, which governs the dynamics of the system. The Hamiltonian is typically chosen to be the sum of the potential energy, defined as the negative logarithm of probability density of the likelihood, and the canonical kinetic energy of the momentum variables.

In each HMC iteration, a proposal is generated by simulating the dynamics of the system for a fixed number of steps using a numerical integration scheme. The acceptance probability of the proposal is then computed using the Metropolis-Hastings (MH) algorithm based on the Hamiltonian energy of the sample. Then, it can be shown that the target distribution can be returned by marginalizing the momenta variables:
\begin{equation} \label{Eq:canonical_target}
    \mathcal{P}(\boldsymbol{x}) \propto \int \exp(-H(\boldsymbol{x}, \boldsymbol{p})) \, d\boldsymbol{p} \, . 
\end{equation}

One of the challenges of HMC is choosing the step size and number of integration steps, which can have a significant impact on the performance of the algorithm. To address this challenge, the No-U-Turn Sampler (NUTS)~\citep{hoffman2011nouturn} algorithm was proposed as an extension of HMC. NUTS addresses this issue by introducing a recursive algorithm that determines the optimal number of steps and step size for each iteration. The algorithm generates a "tree" of proposals by evolving the Hamiltonian dynamics forwards and backwards in time. The number of steps is then chosen such that the trajectory does not turn on itself. This prevents the sampler from returning to previously explored regions of the parameter space. The algorithm then evaluates the Hamiltonian at each new proposal and computes an acceptance probability using MH. The step size is chosen such that a given target acceptance probability is met.

\subsection{MicroCanonical HMC}
\label{sec:mchmc}
Recently, \cite{robnik2022microcanonical} developed a new family of HMC sampling algorithms based on the Micro-Canonical ensemble, named Micro-Canonical Hamiltonian Monte Carlo (MCHMC). The key feature of MCHMC is that it only makes use of a single Hamiltonian energy level to explore the whole parameter space; \textit{i.e.} the Hamiltonian energy of the system is conserved. Mathematically this means that the target distribution is obtained as:
\begin{equation}
    \mathcal{P}(\boldsymbol{x}) \propto \int \delta(H(\boldsymbol{x}, \boldsymbol{p}) - E) \, d\boldsymbol{p} \, .
    \label{eq:mc_target}
\end{equation}
As shown in \cite{robnik2022microcanonical}, there is a multitude of Hamiltonians that satisfy Eq.~\ref{eq:mc_target}. In this work, we use a \julia{} implementation of the variable mass Hamiltonian as presented in \cite{robnik2022microcanonical}.

MCHMC poses several advantages over traditional HMC. First, it does not require the Metropolis adjustment to return unbiased posteriors as in traditional HMC. Second, the bias on the estimated distributions depends on the Hamiltonian error per dimension while in traditional HMC it depends on the absolute Hamiltonian error. Third, the dynamics of the variable mass Hamiltonian result in the particle moving the slowest in the regions where the likelihood density is the highest. This is the opposite behaviour to traditional HMC where the particle moves the fastest where the likelihood density is the highest. In combination, these three features allow MCHMC to explore high dimensional parameter spaces more efficiently than traditional HMC specially for very large numbers of dimensions, as shown in~\cite{Bayer:2023rmj}.

\subsection{Pathfinder}
\label{sec:vi}
Sometimes exploring the parameter space can become too expensive even when the gradient of the likelihood is known. For this reason, the scientific community has developed a series of posterior approximation methods.

The most popular of such methods is variational
inference (VI)~\citep{ranganath2013black, kucukelbir2016automatic}. The goal of VI is to find a tractable approximate distribution to the posterior. In order to do so, VI uses a family of distributions (known as the variational family) in terms of the original parameters of the problem and a series of variational parameters. VI then finds the variational parameters that minimize the Kullback-Leibler (KL) divergence with respect to the posterior. This results in faithful approximations to the posterior even if the actual posterior falls out of the variational family.

Pathfinder is a VI algorithm that locates approximations to the target density along a minimization path~\citep{zhang2022pathfinder}. Pathfinder uses L-BFGS, a quasi-Newton method that relies on gradients to estimate the Hessian matrix. While moving along the minimization path, samples are drawn from a multivariate normal distribution characterized by the estimated Hessian matrix. Pathfinder outperforms other VI methods in providing quick estimates of the posterior, as shown in~\cite{zhang2022pathfinder} for a series of benchmark problems. 

Moreover, Pathfinder can be run in parallel, greatly improving its estimates of the posterior by combining the approximations of each independent run using Pareto-Smoothed Importance Sampling (PSIS) at a low computational cost~\citep{vehtari2022pareto}. Furthermore, PSIS provides a quantitative metric which can be used to assess the quality of the posterior found by Pathfinder. 


Last but not least, Pathfinder can be used to accelerate MCMC methods, using its draws to start chains, whose initial points are usually drawn from the prior and hence are far from the typical set. Since Pathfinder draws are (very likely to be) on the typical set, this reduces the length of the so called burn-in phase, where chains approach the high mass probability region.\footnote{It is possible to further reduce the burn-in phase using the estimated covariance from Pathfinder as the mass matrix provided to the HMC sampler, but we did not take advantage of this feature in this work.}

\section{Results}
\label{sec:results}
In this section we discuss the main results of this work.

We start by testing the accuracy of our emulators performing the MCMC analysis of three different datasets, the \planck{}~\citep{aghanim_planck_2020}, ACT DR4~\citep{aiola_atacama_2020}, and 2018 SPT-3G~\citep{balkenhol23} datasets. The analyses are performed using both \camb{} and \capse{} and the obtained posteriors are compared in order to assess the emulator accuracy.

\subsection{Datasets}
\label{sec:datasets}
\subsubsection{Planck (2018)} \label{sec:planck} 

In this subsection, we show the constraints obtained using data from the \planck{} satellite. Specifically, we consider the compressed temperature and polarization \planck{} 2018 \texttt{Plik\_lite} likelihood, which is constructed from the high-$\ell$ $TT+TE+EE$ CMB bandpowers and covariances that have been premarginalized over astrophysical contributions in the spectrum estimation step~\citep{Planck:2019nip}.
The only nuisance parameter left in this compressed likelihood is the overall \planck{} calibration factor $A_\mathrm{Planck}$ which simply rescales the foreground-marginalized bandpowers as $\hat{C}_{\ell}\to\hat{C}_{\ell}/A_\mathrm{Planck}^2$. 
This likelihood is available in \texttt{Cobaya} as \texttt{planck\_2018\_highl\_plik.TTTEEE\_lite}.

\begin{figure*}%
\centering%
\includegraphics[width=\textwidth]{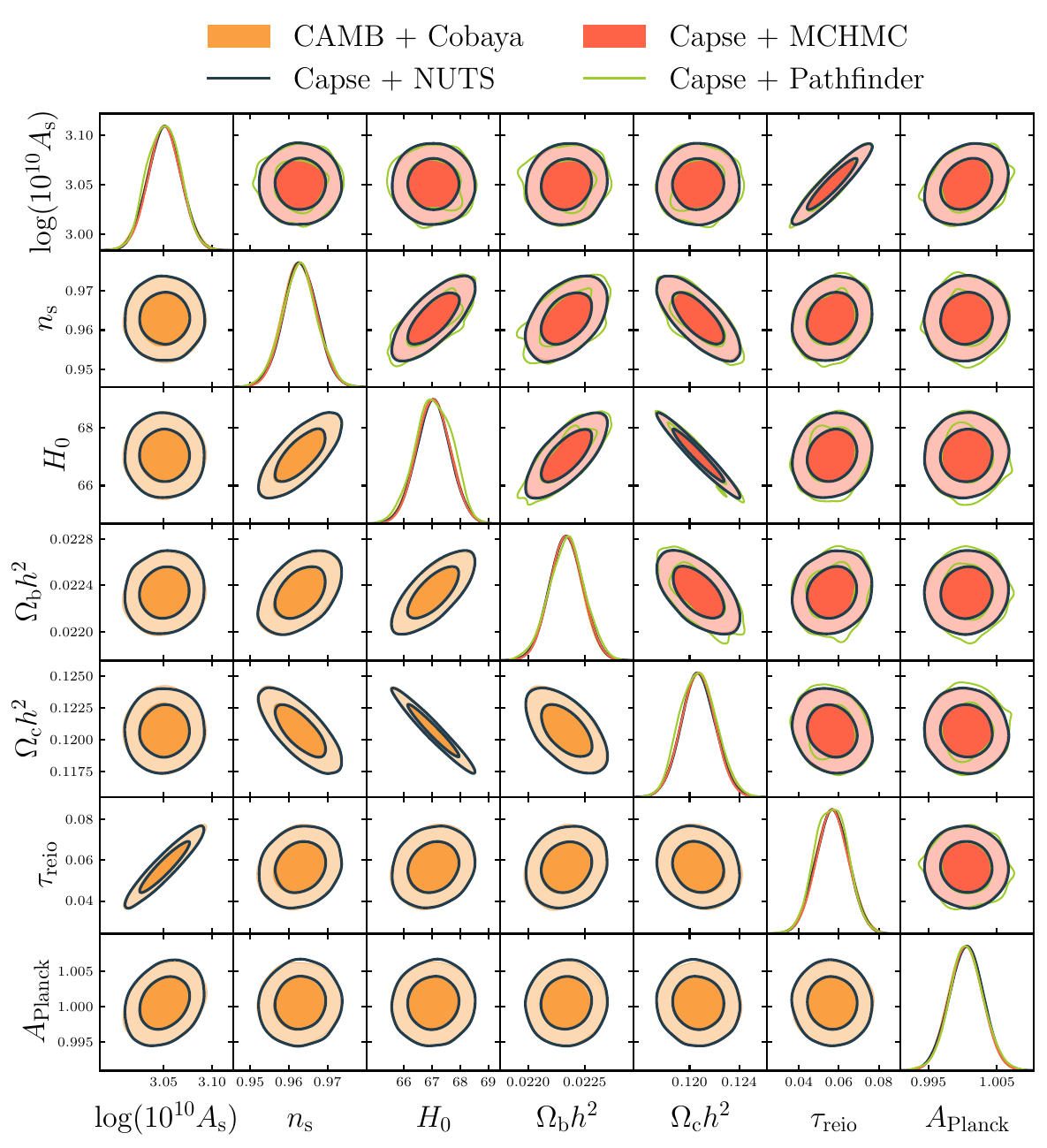}%
\caption{%
Triangle plot, showing the standard \planck{} lite chains and the one obtained using \capse{} in combination with \turing{}.
The lower triangular part of the plot focuses on the comparison between \camb{} and \capse{}, in order to ensure the precision of our emulators and likelihoods. The upper half of the plot shows only \capse{} contours, with a comparison of the results between the samplers employed in our analysis, namely NUTS, MCHMC, and Pathfinder.
}%
\label{fig:contour_planck}
\end{figure*}

In Fig.~\ref{fig:contour_planck}, we compare the contours obtained with \camb{} and \capse{}. The former has been used within \cobaya{} and a nuisance-marginalized \planck{} likelihood, with a termination criterion of $|\hat{R}-1|<0.01$, requiring 6 hours to converge with 4 parallel chains, each of them with 4 threads, for a total of around 100 CPU-hours. The latter employed \plancklite{},\footnote{\url{https://github.com/JuliaCosmologicalLikelihoods/PlanckLite.jl}} a pure \julia{} implementation of the \planck{} lite likelihood, and used 6 NUTS chains, with 500 burn-in and 1,000 accepted steps, on a laptop in around 4 minutes of wall-clock time using less than $0.5$ CPU-hours, but still reaching an $\hat{R}-1$ smaller than 0.002 for all parameters. We emphasize that the NUTS analyses was performed with only a handful of parameters to tune: the number of parallel chains, the burn-in and sampling steps per chains, and the target acceptance ratio. We did not provide any input such as a covariance matrix or a proposal step-size for each parameter.
This shows the robustness of the NUTS algorithm, which is able to perform a Bayesian analysis with little to no input from the user. This is very different from the MH sampler implemented in \cobaya{}, which greatly benefits in its performance from being finely tuned by the user.

Regarding the accuracy of the analysis, the two contours are very similar, with differences on the marginalized 1D posteriors smaller than $0.1\,\sigma$, as is shown in the lower half of Fig.~\ref{fig:contour_planck}.

\subsubsection{ACT DR4} \label{sec:ACT}
\begin{figure*}%
\centering%
\includegraphics[width=\textwidth]{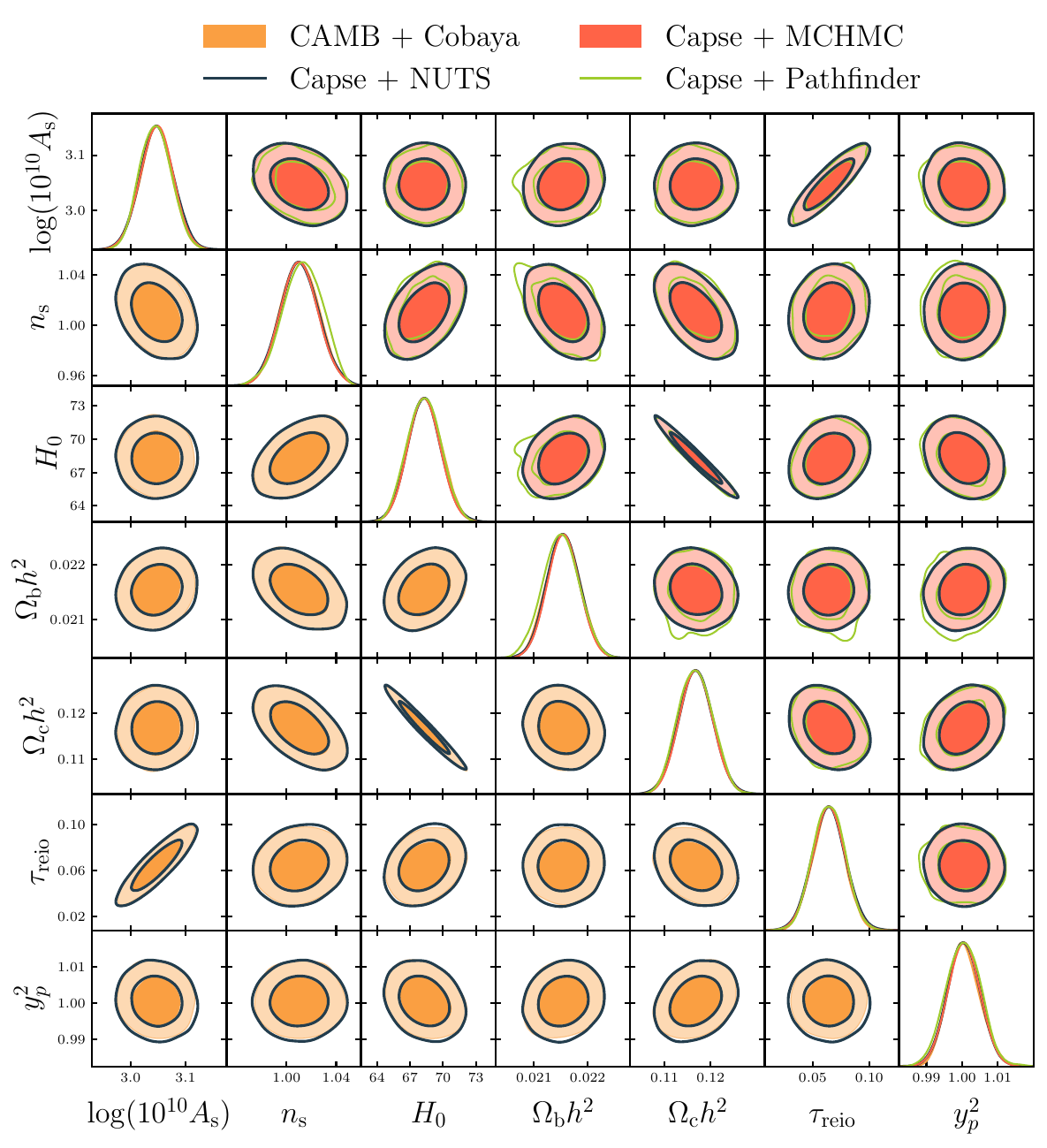}%
\caption{%
Triangle plot, showing the standard ACTPolLite chains and the one obtained using \capse{} in combination with \turing{}.
The lower triangular part of the plot focuses on the comparison between \camb{} and \capse{}, in order to ensure the precision of our emulators and likelihoods. The upper half of the plot shows only \capse{} contours, with a comparison of the results between the samplers employed in our analysis, namely NUTS, MCHMC, and Pathfinder.}%
\label{fig:contour_act}
\end{figure*}

Here we show the results from the chains based on the ACT  Data Release 4 (DR4) likelihood \citep{aiola_atacama_2020}.\footnote{\url{https://github.com/ACTCollaboration/pyactlike}}
This likelihood is constructed from the temperature and polarization $TT/TE/EE$ power spectra measured with ACT data taken between 2013 and 2016 covering more than 17,000 deg$^2$ of the sky with spatially varying depth.
Specifically, the ACT likelihood relies on cleaned CMB bandpowers in the multipole range $300 \lesssim \ell \lesssim 4000$ that have been marginalized over foreground emissions including the Sunyaev-Zel'dovich effects, diffuse Galactic emission, and the cosmic infrared background.
The covariance of the CMB power spectra already includes contributions from foreground, calibration, and beam uncertainties.
The only nuisance parameter that is allowed to freely vary is the overall polarization efficiency $y_p$, which linearly and quadratically rescales the $C_{\ell}^{TE}$ and $C_{\ell}^{EE}$ spectra, respectively. Also in this case we have coded a pure \julia{} version of this likelihood, \texttt{ACTPolLite.jl}.\footnote{\url{https://github.com/JuliaCosmologicalLikelihoods/ACTPolLite.jl}}
For the \camb{}$+$\cobaya{} chains we used the same hardware settings from before. However, since we used the high-precision settings for \camb{}, the computational resources employed are higher than before, being around 480 CPU-hours required to reach convergence. Regarding our \capse{}$+$\turing{}, we run as before 6 chains in parallel with 500 burn-in steps and 1,000 accepted steps, obtaining convergent chains with $|\hat{R}-1|<0.002$ using around one CPU-hour. Also in this case the agreement between the \camb{} and the \capse{} chains is excellent, with differences on the posteriors smaller than $0.1\,\sigma$, as shown in the lower half of Fig.~\ref{fig:contour_act}.

\subsubsection{SPT-3G}\label{sec:SPT}
\begin{figure*}%
\centering%
\includegraphics[width=\textwidth]{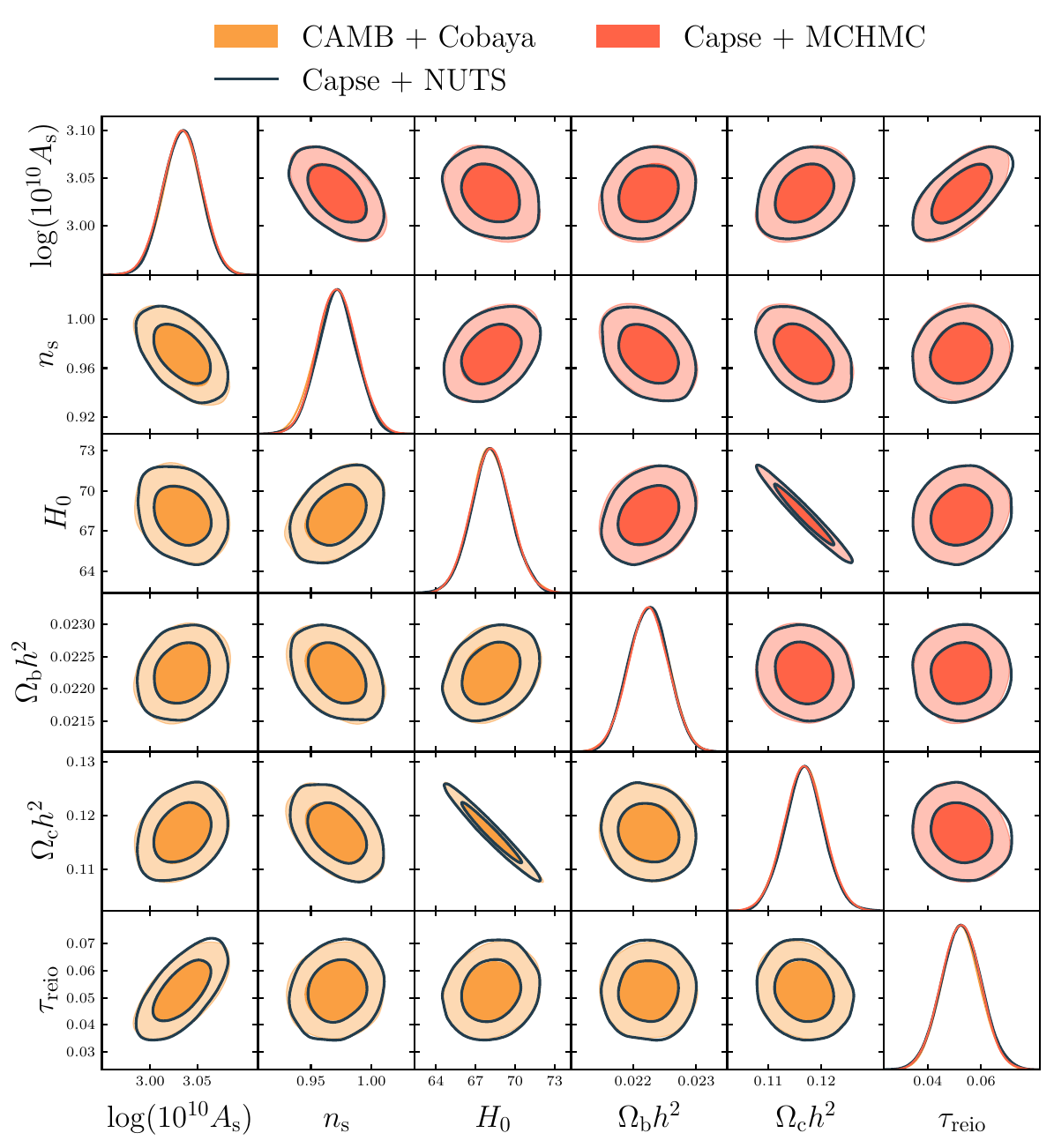}%
\caption{%
Triangle plot, showing the standard SPT-3G chains and the one obtained using \capse{} in combination with \turing{}.
The lower triangular part of the plot focuses on the comparison between \camb{} and \capse{}, in order to ensure the precision of our emulators and likelihoods. The upper half of the plot shows only \capse{} contours, with a comparison of the results between the samplers employed in our analysis, namely NUTS and MCHMC.}%
\label{fig:contour_spt}
\end{figure*}

The final CMB likelihood considered in this study pertains to the multifrequency 2018 SPT-3G temperature and polarization dataset, as detailed in \citet{balkenhol23}.\footnote{\url{https://pole.uchicago.edu/public/data/balkenhol22/}}
This dataset comprises measurements of the CMB $TT/TE/EE$ power spectra across $\sim\,$1,500 deg$^2$ of the southern Sky.
Specifically, the $TT$ power spectra span the angular multipole range from $750 < \ell \le 3,000$,while the $TE$ and $EE$ spectra extend from $300 < \ell \le 3,000$.
Distinct from the \texttt{PlancklLite.jl} and \texttt{ACTPolLite.jl} likelihoods, which are premarginalized over astrophysical emissions, the SPT-3G one employs a parametric model for foreground characterization in both temperature and polarization.
Additionally, the SPT-3G likelihood accounts for the effects of instrumental calibration and beam uncertainties, aberration due to the relative motion with respect to the CMB rest frame, and super-sample lensing \citep{dutcher21}.
Owing to its large number of foreground and nuisance parameters, the SPT-3G likelihood serves as an exemplary testbed for gradient-based sampling techniques. 
More precisely, in addition to the 6 $\Lambda$CDM parameters, we sample over 6 calibration parameters, 26 foreground parameters, and a nuisance parameter (super-sample lensing convergence), for a total of 39 parameters.
A thorough description of the data model can be found in \cite{balkenhol23}.
A pure \julia{} implementation of the SPT-3G likelihood, \texttt{SPTLikelihoods.jl}, is made available in this case too.\footnote{\url{https://github.com/JuliaCosmologicalLikelihoods/SPTLikelihoods.jl}}
For the \camb{}$+$\cobaya{} chains we used the same hardware settings from before. However, since we used the high-precision settings for \camb{}, the computational resources employed are higher than before, being around 1,600 CPU-hours required to reach convergence. We also run 6 chains, with 500 burn-in steps and 1,000 accepted steps, using \capse{}$+$\turing{}: the obtained chains have a convergence with $|\hat{R}-1|<0.002$ using 14 CPU hours. Also with this more challenging posterior there is an excellent agreement between the standard and the emulator chains, with differences on the posterior smaller than $0.1\,\sigma$, as shown in Fig.~\ref{fig:contour_spt}.

\subsection{Samplers comparison}
\label{sec:samplers_comparison}

\begin{table*} 
\centering
\begin{tabular}{|c|c|c|c|c|}
\hline
 & \camb{}+\cobaya{} & \capse{}+NUTS & \capse{}+MCHMC & \capse{}+Pathfinder  \\ 
\hline
\multicolumn{5}{|c|}{\planck{}, Parameter and 95\% limits} \\
\hline
$\ln 10^{10}A_s$ & $3.051^{+0.035}_{-0.035}$ & $3.051^{+0.033}_{-0.033}$  & $3.051^{+0.033}_{-0.033}$  & $3.050^{+0.031}_{-0.031}$ \\
$n_s$  & $0.9626^{+0.0089}_{-0.0087}$  & $0.9630^{+0.0085}_{-0.0087}$  & $0.9630^{+0.0085}_{-0.0084}$  & $0.9628^{+0.0092}_{-0.0090}$ \\
$H_0[\mathrm{km}/s/\mathrm{Mpc}]$  & $67.0^{+1.2}_{-1.2}$  & $67.0^{+1.2}_{-1.2}$  & $67.1^{+1.2}_{-1.2}$  & $67.1^{+1.2}_{-1.3}$\\
$\omega_b$  & $0.02233^{+0.00029}_{-0.00029}$  & $0.02234^{+0.00029}_{-0.00029}$  & $0.02234^{+0.00029}_{-0.00028}$  & $0.02234^{+0.00029}_{-0.00030}$ \\
$\omega_c$  &$0.1208^{+0.0028}_{-0.0028}$  & $0.1207^{+0.0027}_{-0.0027}$ & $0.1207^{+0.0027}_{-0.0026}$  & $0.1207^{+0.0029}_{-0.0028}$ \\
$\tau$ & $0.057^{+0.017}_{-0.016}$ & $0.057^{+0.016}_{-0.016}$ & $0.057^{+0.016}_{-0.016}$ & $0.056^{+0.015}_{-0.015}$ \\
$A_\textrm{Planck}$ & $1.0004^{+0.0047}_{-0.0048}$  & $1.0005^{+0.0049}_{-0.0049}$ & $1.0005^{+0.0048}_{-0.0047}$  & $1.0006^{+0.0047}_{-0.0046}$ \\
\hline

\multicolumn{5}{|c|}{ACT, Parameter means and 95\% limits} \\
\hline
$\ln 10^{10}A_s$  & $3.047^{+0.056}_{-0.057}$  & $3.048^{+0.060}_{-0.059}$  & $3.047^{+0.058}_{-0.059}$  & $3.045^{+0.053}_{-0.050}$ \\
$n_s$  & $1.011^{+0.029}_{-0.028}$  & $1.010^{+0.031}_{-0.030}$  & $1.010^{+0.030}_{-0.029}$  & $1.013^{+0.031}_{-0.030}$ \\
$H_0[\mathrm{km}/s/\mathrm{Mpc}]$  & $68.4^{+3.0}_{-2.9}$  & $68.3^{+3.0}_{-2.9}$  & $68.3^{+2.9}_{-2.9}$  & $68.3^{+2.7}_{-2.6}$\\
$\omega_b$  & $0.02155^{+0.00060}_{-0.00058}$  & $0.02155^{+0.00061}_{-0.00060}$  & $0.02155^{+0.00059}_{-0.00059}$ & $0.02150^{+0.00061}_{-0.00061}$\\
$\omega_c$  & $0.1167^{+0.0071}_{-0.0073}$  & $0.1168^{+0.0073}_{-0.0072}$  & $0.1168^{+0.0072}_{-0.0070}$  & $0.1167^{+0.0066}_{-0.0063}$ \\
$\tau$ & $0.064^{+0.026}_{-0.027}$ & $0.064^{+0.029}_{-0.028}$ & $0.064^{+0.028}_{-0.028}   $ & $0.064^{+0.025}_{-0.025}$ \\
$y_p$  & $1.0006^{+0.0095}_{-0.0084}$  & $1.0005^{+0.0090}_{-0.0091}$  & $1.0005^{+0.0089}_{-0.0090}$  & $1.0004^{+0.0091}_{-0.0091}$ \\
\hline

\multicolumn{5}{|c|}{SPT, Parameter means and 95\% limits} \\
\hline
$\ln 10^{10}A_s$  & $3.034^{+0.041}_{-0.041}$ & $3.034^{+0.039}_{-0.038}$  & $3.034^{+0.040}_{-0.041}$  & $-$ \\
$n_s$  & $0.971^{+0.032}_{-0.033}$  & $0.972^{+0.030}_{-0.031}$  & $0.972^{+0.033}_{-0.032}$  & $-$ \\
$H_0[\mathrm{km}/s/\mathrm{Mpc}]$  & $68.2^{+3.1}_{-2.9}$  & $68.2^{+3.0}_{-2.9}$  & $68.2^{+3.0}_{-2.9}$  & $-$\\
$\omega_b$  & $0.02225^{+0.00061}_{-0.00062}$  & $0.02225^{+0.00061}_{-0.00059}$  & $0.02224^{+0.00062}_{-0.00062}$ & $-$\\
$\omega_c$  & $0.1168^{+0.0074}_{-0.0074}$  & $0.1168^{+0.0073}_{-0.0073}$  & $0.1168^{+0.0075}_{-0.0074}$  & $-$ \\
$\tau$ & $0.052^{+0.015}_{-0.014}$ & $0.053^{+0.015}_{-0.015}$ & $0.052^{+0.014}_{-0.015}$ & $-$ \\
\hline
\end{tabular}
\caption{Mean and 95\% limits of the 1D marginal posteriors of the cosmological parameters obtained from the \planck{}, ACT, and SPT analyses, respectively. Different columns show the constraints obtained performing the same analyses using the sampling algorithms MH, NUTS, MCHMC and Pathfinder. The mean and limits have been computed using \texttt{GetDist}. Given the highly unstable results obtained when applying Pathfinder to the SPT posterior, we did not quote any number for this specific application.}
\label{tab:constraints}
\end{table*}

\begin{table*} 
\centering
\begin{tabular}{|c|c|c|c|c|c|c|}
\hline
& \multicolumn{2}{c}{\planck{}} & \multicolumn{2}{|c|}{ACT} & \multicolumn{2}{|c|}{SPT}\\
\hline
& ESS/s & Comp. (s) & ESS/s & Comp. (s) & ESS/s & Comp. (s) \\
\hline
\camb{}+\cobaya{} & 0.004  & 345,600 & 0.0003 & 1,728,000 & 0.0003 & 5,760,000 \\
\capse{}+NUTS  & 1.597  & 7,014  & 1.914  & 11,813 & 0.09 & 50,400\\
\capse{}+MCHMC  & 3.686  & 4,374  & 4.136  & 7,792 & 0.335 & 43,200\\
\capse{}+Pathfinder  & N/A  & 15 & N/A  & 16 & - & -\\
\hline
\end{tabular}
\caption{Shows the ESS per second and computation time achieved by the \cobaya{}, NUTS, MCHMC and Pathfinder inference algorithms when performing the \planck{} (2nd and 3rd columns),  ACT (4th and 5th columns), and SPT (6th and 7th column) analyses. Note that \cobaya{}  used the cosmological code \camb{} while the other samplers were paired with the \capse{} emulator. Given the highly unstable results obtained when applying Pathfinder to the SPT posterior, we did not quote any number for this specific application.}
\label{tab:performance}
\end{table*}

After validating our emulators, we now focus on the different inference algorithms and study their impact on the speed, accuracy, and robustness of the analyses.

The compatibility of the \capse{} emulating framework with AD methods makes it extremely synergistic with a plethora of gradient-based samplers. Here we compare the performance of three gradient-based samplers (see Sec.~\ref{sec:samplers} for details). First, we compare the state-of-the-art NUTS sampler used to obtain our fiducial results in Sec.~\ref{sec:datasets} with the recently developed MCHMC sampler. Second, we compare NUTS against a VI method, Pathfinder\footnote{We use the publicly available \julia{} implementations of \href{https://github.com/TuringLang/AdvancedHMC.jl}{NUTS}, \href{https://github.com/JaimeRZP/MicroCanonicalHMC.jl}{\texttt{MicroCanonicalHMC.jl}} and \href{https://github.com/mlcolab/Pathfinder.jl}{\texttt{Pathfinder.jl}}.}. In all cases we use the same likelihoods described in Sec.~\ref{sec:datasets} to draw the comparisons. This is possible thanks to the modularity of the \julia{} ecosystem, which allows us to write a likelihood in \turing{} and then sample it using different libraries and algorithms.

\subsubsection{NUTS vs MCHMC}
\label{sec:nuts_vs_mchmc}
Before comparing the performance of the two samplers, it is essential to verify that NUTS and MCHMC return the same posterior distributions. As MCMC methods, both NUTS and MCHMC should asymptotically converge to the same target distribution.  The comparison of the NUTS and MCHMC posteriors for \planck{}, ACT, and SPT analyses can be found in Figs.~\ref{fig:contour_planck}, respectively. The corresponding numerical values for the quoted constraints can be found in Tab.~\ref{tab:constraints}. In this table, we observe that the constraints obtained from both NUTS and MCHMC are within a 0.1$\sigma$ distance of their Metropolis-Hastings (MH) counterpart. The excellent agreement between the two samplers indicates that differences are well within the sampling noise.

Having established that both samplers converge to the same posterior, the pivotal question now is how quickly they achieve convergence. To address this, we assess the effective sample size (ESS) per second, as it provides insight into the convergence rate of MCMC chains~\citep{brooks2011handbook}. The auto-correlation among samples in an MCMC chain introduces uncertainty in estimating the mean and standard deviation of the posterior distributions. A higher number of effective samples per second indicates quicker convergence.

We compute the ESS of all chains using the \texttt{Python} library \texttt{GetDist}. When analyzing \planck{}, we find an ESS/s of 1.6 for the NUTS sampler, while MCHMC returns an ESS/s of 3.7, as can be seen in Tab.~\ref{tab:performance}. In the case of ACT, NUTS achieves 1.9 ESS/s while MCHMC reaches 4.1. Thus in both analyses we see that MCHMC is twice as fast as NUTS, as expected on the findings of  \cite{robnik2022microcanonical}. The difference is even bigger for the SPT analysis, since NUTS and MCHMC delivers an ESS/s of, respectively, 0.09 and 0.335\footnote{The overall decreased performance for the SPT analysis, compared to the \planck{} and ACT ones, is mostly driven by the more expensive likelihood, since more operations are needed to go from the input $C_\ell$ spectra to the log-likelihood calculation.}. Overall, our findings show that the MCHMC sampler has an higher efficiency compared to the NUTS for the applications considered, as in~\citep{Bayer:2023rmj}.\footnote{We would like to emphasize that \texttt{MicroCanonicalHMC.jl} is still an alpha-software which is being actively developed: performance might improve with a more stable and refined implementation of MCHMC within this library.}

\subsubsection{NUTS vs Pathfinder}
\label{sec:nuts_vs_pathfinder}
Comparing NUTS to Pathfinder is a more difficult task since Pathfinder is a VI method while NUTS is an MCMC method. VI methods do not guarantee asymptotic convergence to the target distribution, unlike MCMC methods. Instead, VI approximates the posterior distribution as a combination of possible distributions (see Sec.~\ref{sec:samplers} for details). Moreover, the concept of ESS is not applicable to VI methods since they do not inherently produce MCMC chains, although MCMC-like samples can be generated by sampling the approximated posterior.

Despite these caveats, we observe excellent agreement between the posteriors obtained with Pathfinder and NUTS for all the likelihoods considered. 
This can be observed in the top half of Figs.~\ref{fig:contour_planck} and ~\ref{fig:contour_act}. In these plots it is possible to see that Pathfinder reproduces not only the mean and standard deviation of the marginal posteriors but also the different correlations between the parameters. Looking at Tab.~\ref{tab:constraints} we can see that the Pathfinder constraints are once again within $0.2\,\sigma$ of the \camb{}+MH constraints.\footnote{These results have been generated using the ensemble version of Pathfinder, multi Pathfinder, that runs several VI processes in parallel to produce better approximations; in particular, for each multi Pathfinder realization, we run 6 single Pathfinder realization in parallel.}

To assess the reliability of Pathfinder, we computed 1,000 realizations of it and found that the mean obtained for each parameter follows a distribution which is close to a Gaussian with a standard deviation corresponding to less than $0.1,\sigma$, measured on each parameter when performing a full MCMC analysis.

Regarding the SPT analysis, we found Pathfinder to provide unstable results when applied to this likelihood. For this reason, we decided not to show any result related to this combination of likelihood and sampler.

Regarding performance, since the ESS is not well-defined for VI methods, we compare the speed of the two methods by examining pure wall times. As shown in Tab.~\ref{tab:performance}, Pathfinder successfully approximates the posteriors of the \planck{} and ACT analyses in around 10 seconds. This represents a 5 to 6 orders of magnitude improvement compared to the \cobaya{} analyses using \camb{} and a 2 orders of magnitude improvement compared to using \capse{} with either NUTS or MCHMC. However, we emphasize that the robustness of variational methods can significantly decrease when the posterior is everything but a multivariate normal distribution. However, even in these cases, methods such as Pathfinder still can prove useful as they can be used to initialize standard MCMCs, reducing the burn-in phase even when considering challenging posteriors with funnel-like geometries~\citep{zhang2022pathfinder}. This is actually what we did for the SPT analysis: the Pathfinder draws were unstable and likely to be wrong, as detected by the PSIS. Nevertheless, they were way closer to the typical set than random draws from the prior and hence used them to initialize our NUTS and MCHMC chains.

\subsection{MOPED compression \& Chebyshev decomposition}
All the MCMC results shown in the previous section are based on the standard emulator and likelihood. Although the favourable comparison with \camb{}$+$\cobaya{} approach, there is still room for some computational improvements. In this section we show results based on \plancklite{} and on a \texttt{MOPED} compressed likelihood and the Chebyshev-based emulators.

We begin by discussing the standard likelihood results. As discussed in Sec.~\ref{sec:architecture}, we exploit the Chebyshev decomposition to achieve peak performance. The \plancklite{} likelihood is composed of a sequence of linear operators that remain independent of cosmological parameters. Utilizing this property, we can apply these operators to the Chebyshev basis functions and precompute and store the results. During each step of the MCMC, we only need to multiply the stored quantities by the expansion coefficients, which solely contain the cosmological dependence. This approach eliminates the need to repeatedly apply the linear operator, resulting in enhanced computational efficiency.

Traditionally, optimizing the computation of the likelihood is not a priority since the bottleneck is the computation of the theory prediction. However, this is no longer the case when emulators are employed. For \capse{},  the $C_\ell$'s computation is around 7 times faster than the computations required to compute the log-likelihood such as the $C\ell$'s binning. This highlights the performance of our emulators, since the bottleneck is now represented by the likelihood itself. When using NUTS and MCHMC we find an ESS/s of 13 and 32 respectively. Meanwhile, the computation time when using Pathfinder is reduced to 1.5 seconds. This shows an almost uniform performance improvement by a factor of 8 across the different inference methods considered.

Finally, we employ a \texttt{MOPED} likelihood~\citep{prince_data_2019}, a data compression technique which reduces the dimensions of the data vector to one number per parameter of interest, requiring an even lower amount of computational resources. We find an exquisite agreement between the \plancklite{} and the \texttt{MOPED}  posteriors, with differences on marginalized parameters smaller than $0.1\,\sigma$ as shown in Fig.\ref{fig:planck_MOPED}, but with an improved computational performance: NUTS and MCHMC reach an ESS/s of, respectively, 60 and 150, while Pathfinder runs in 0.5 seconds. Compared to the performance obtained when using the standard emulator and \plancklite{} we report an improvement of a factor of 40-50. This brings down the wall time for the \planck{} analysis to a record one second. The finding of this section are summarized in Table~\ref{tab:performance_cheb}.

\begin{table} 
\centering
\begin{tabular}{|c|c|c|}
\hline
& \multicolumn{2}{c|}{\planck{}}\\
\hline
& ESS/s & Comp. (s)\\
\hline
\plancklite{}+NUTS  & 13  & 1031\\
MOPED+NUTS  & 60  & 228\\
\plancklite{}+MCHMC  & 32  & 605\\
MOPED+MCHMC  & 150 & 157\\
\plancklite{}+Pathfinder  & N/A  & 1.5\\
MOPED+Pathfinder  & N/A  & 0.5\\
\hline
\end{tabular}
\caption{Shows the ESS per second and computation time achieved by the NUTS, MCHMC and Pathfinder inference algorithms when performing the \planck{} Analysis. Note that here we are using the Chebyshev based emulator.}
\label{tab:performance_cheb}
\end{table}

\section{Comparison with other emulators}
\label{sec:comparison}
In this section we compare the emulators presented in this work to two closely related alternatives in the literature, \classnet{}~\citep{albers_cosmicnet_2019} and the recently developed \cosmopower{}\footnote{Unless otherwise stated, in this section we consider the \cosmopower{} implementation of~\cite{bolliet_high-accuracy_2023}.}~\citep{mancini_itcosmopower_2021, bolliet_high-accuracy_2023}. \classnet{} emulates the source functions required to compute the CMB angular power spectra using \texttt{Python}-based neural network emulators. The result of the emulation is then passed to \texttt{CLASS}, which computes the line-of-sight integrals. The main advantage of this approach is that \classnet{} does not need to include $A_s$ and $n_s$ in the emulation space, since the primordial power spectrum is analytically computed and integrated against the emulated source functions. \cosmopower{} is a  \texttt{TensorFlow} based emulator that follows a design philosophy closer to \capse{}, as it directly emulates the CMB angular power spectra. Initially, it focused on the \planck{} analysis  but  it was later extended to $\ell_\mathrm{max}=10,000$ and the same high-accuracy settings employed here~\citep{bolliet_high-accuracy_2023}.

Comparing emulators is challenging due to their distinct architectures and optimization goals. Here we highlight the differences between  \classnet{}, \cosmopower{} and \capse{}, explaining the pros and cons of each framework.

We decided not to perform a comparison with other frameworks, such as those presented in~\cite{nygaard_connect_2022, Gunther:2023xhh, To:2022ubu} as they have a different goal: performing an analysis using the lowest possible number of Boltzmann code evaluations using active learning schemes.

In order to compare \capse{}, \classnet{}, and \cosmopower{} we will use several metrics. First, we will compare the computational performance of each emulator. Second, we will compare the training efficiency. Third, we will look at the flexibility of the emulator, \textit{i.e.} how often would it need to be retrained based on the changes to the analysis that one wishes to perform. Finally, we will also study the efficiency of the sampling algorithms compatible with each emulator.

We report the speedup obtained by the three emulation frameworks when compared to a traditional Boltzmann solver in the first row of Tab.~\ref{tab:comp}. In this table, we can see that \classnet{}, \cosmopower{}, and \capse{} emulators report a speedup of a factor of 3,\footnote{We report that in \cite{gunther_cosmicnet_2022} the authors performed a \planck{} data analysis, without focusing on high-precision settings in which they would likely achieve a higher speed-up.} 1,000, and 1,000,000 respectively. We would like to emphasize that the numbers shown in the comparison were taken directly from the \classnet{} and \cosmopower{} release papers; their performance might improve with an hyperparameter optimization.

\begin{table*} 
\centering
\begin{tabular}{|c|c|c|c|}
\hline
& \classnet{}  & \cosmopower{} & \capse{} \\
\hline
Speedup           & 3 & 1,000 & 1,000,000 \\
Architecture      & NN$^*$ & 4x512-NN + PCA & 5x64-NN + Chebyshev \\
Training points   & $10^4$ & $10^5$ & $10^4$ \\
Tessellation      & Hyperellipsoid & LHC & LHC \\
\hline
\multicolumn{4}{|c|}{Emulation domain} \\
\hline
 $\ln 10^{10} A_{\mathrm{s}}$          & N/A        & [2.5, 2.5] & [2.5, 3.5] \\
$n_{\mathrm{s}}$                  & N/A        & [0.8812, 1.0492] & [0.88, 1.06] \\
$\tau$                            & $N(0.0561, 0.0071)$ & [0.02, 0.12] & [0.02, 0.12] \\
$H_0[\mathrm{km}/s/\mathrm{Mpc}]$ & $N(67.66, 0.42)$ & [39.99, 100.01] & [40, 100] \\
$\omega_{\mathrm{b}}$             & $N(0.02242, 0.00014)$ & [0.0193, 0.02533] & [0.0193, 0.02533] \\
$\omega_{\mathrm{c}}$             & $N(0.11933, 0.00091)$ & [0.08, 0.2] & [0.08, 0.2] \\
$\ell_\mathrm{max}$ & 2,500 & 11,000& 5,000\\ 
\hline
\end{tabular}
\caption{A comparison between the emulating frameworks \classnet{}, \cosmopower{}, and \capse{}. Note that the architecture of \classnet{} involves 7 different NN with 3 to 4 layers each adding to a total 5366 neurons. Regarding \cosmopower{}, here we refer to the emulator presented in~\cite{bolliet_high-accuracy_2023}.}
\label{tab:comp}
\end{table*}

Comparing the training efficiency of the three emulators is not straightforward, primarily due to two main differences between them. Firstly, the number of inputs (\textit{i.e.} the cosmological parameters) and their ranges vary from emulator to emulator. While \cosmopower{} and \capse{} were trained on the same number of input features, \classnet{} requires two fewer inputs, significantly reducing the volume of training data needed to cover its domain. Secondly, the tessellation method to cover the parameter space can also differ. \cosmopower{} and \capse{} use a Latin hypercube to span their domains, while \classnet{} was trained on a hyperellipsoid centered around the best-fit \planck{} TT,TE,EE+lowE+lensing+BAO $\Lambda$CDM cosmology~\citep{aghanim_planck_2020}, spanning 6 $\sigma$'s.
Regarding the number of learned features, \classnet{} learns source functions on a $k$-grid with $\mathcal{O}(700)$ elements, chosen to accurately compute the \planck{} $C_\ell$'s. 
In contrast, \capse{} computes all $C_\ell$'s up to $\ell_\mathrm{max}=5,000$ while \cosmopower{} reaches an $\ell_\mathrm{max}$ of 11,000, making them both suitable for next-generation CMB experiments.

Considering the aforementioned caveats, training \capse{} requires $10^4$ training points, which is of the same order of magnitude as \classnet{}, and an order of magnitude less than \cosmopower{}~\cite{bolliet_high-accuracy_2023}. In terms of network architecture, \cosmopower{} employs a one-size-fits-all approach with 4 fully connected hidden layers, each containing 512 neurons. \classnet{}, on the other hand, tailors its networks for each emulator, utilizing different architectures with 2-4 hidden layers and 100-500 neurons. \capse{} employs a consistent architecture across all networks, consisting of 5 hidden layers with 64 neurons. In terms of training time, \capse{} takes slightly less than one hour using a CPU, while the training of \cosmopower{} lasted one hour using a GPU. After training, \capse{} is slightly more precise than \cosmopower{} for the $\phi\phi$ coefficients, while \cosmopower{} is 2-3 times more precise for the $TT$, $TE$, $EE$ coefficients.

Regarding flexibility, \classnet{} is the most versatile of the emulators. As previously described, it only emulates the transfer functions and computes the line-of-sight integrals using \texttt{CLASS}. 
This allows the user to change the computation of the primordial power spectrum or the method to compute the line-of-sight integrals without having to retrain the emulator, as it would be the case for \capse{} and \cosmopower{}. 

Finally, we studied the efficiency of sampling algorithms compatible with each emulator. The wall time users will experience depends on both the speed of the emulator and the sampling algorithms available for Bayesian inference.  \classnet{} employs traditional gradient-free sampling methods while \capse{} and \cosmopower{} are compatible with state-of-the-art gradient-based algorithms. The original \cosmopower{}  was written in \texttt{TensorFlow} and, coupled to \texttt{TensorFlowProbability}\footnote{\href{https://www.tensorflow.org/probability}{https://www.tensorflow.org/probability}}, has in principle access to a rich variety of gradient-based samplers. Moreover, \cosmopower{} was recently re-written in \texttt{JAX}~\citep{Piras_2023} and embedded in the \texttt{JAX-COSMO} framework, explicitly designed for gradient-based sampling methods.
On the other hand, using these methods with \classnet{} would be technically challenging due to its mixed \texttt{C} and \texttt{Python} implementation.  \capse{} was coupled with the \turing{} PPL, enabling the use of NUTS, MCHMC, and Pathfinder inference algorithms, fully leveraging the differentiability of our emulators.

When comparing the wall time required to compute converging chains, \classnet{} reports a speedup by a factor of 3. In \cite{mancini_itcosmopower_2021}, a \planck{} analysis was performed in $\mathcal{O}(10\,s)$ using a GPU; however, as can be seen on the official notebook prepared by the authors, the chains were initialized on the typical set, using the knowledge coming from previous analyses. \capse{} is able to perform a \planck{} analysis in around a minute with MCMC methods or $\mathcal{O}(1\,s)$ with variational methods. We emphasize that \capse{}'s runs were performed randomly initializing our algorithms from the prior. While there is nothing conceptually wrong in initializing the chains on the typical set, we do think that starting using random draws from the prior is more representative of what happens when analyzing a new dataset for the first time.

Finally, we want to emphasize that the results shown in this comparison are a direct consequence of the design choices of the emulators. \classnet{} is the most flexible framework, but this comes at the expense of a lower speedup. \cosmopower{} delivers a higher speedup than \classnet{} and has already been trained to analyze next-generation CMB surveys, but its training requires an expensive NN architecture and training dataset. \capse{} is faster than \cosmopower{}, can be used to analyze actual datasets and is cheaper to train, but its multipole range and accuracy are slightly lower than the one of \cosmopower{}. According to the scenario considered, one of the three emulators considered will better suit the requirements of the analysis. We postpone to a further publication a detailed investigation of the origin of the higher runtime performance of \capse{}.

\section{Conclusions}
\label{sec:conclusions}
In this paper, we introduced \capse{}, a novel emulator framework for CMB angular power spectra. We discussed the preprocessing steps and the characteristics of the NN library employed, highlighting the distinctions from similar emulators in the literature~\citep{mancini_itcosmopower_2021, gunther_cosmicnet_2022}.
We additionally developed an emulator based on Chebyshev polynomial decomposition, which significantly enhanced the computational performance of the original emulator by reducing the number of output features. Leveraging the Chebyshev linear decomposition further improved the efficiency of the log-likelihood evaluation.

To validate the accuracy of our emulators, we conducted two tests. First, we compared \capse{} and \camb{} computations over 20,000 combinations of input parameters across the entire emulation range. Our emulators exhibited errors below $0.1\,\sigma$ for all scales relevant to future CMB surveys. Second, we analyzed the \planck{} and ACT DR4 nuisance-marginalized datasets as well as the full 2018 SPT-3G dataset, comparing the posteriors with those obtained using \camb{} and \cobaya{}. The derived contours showed a remarkable agreement, better than $0.1\,\sigma$.

Next, we examined the performance impact of different sampling algorithms. 
For this purpose, we considered several gradient-based samplers, namely NUTS, the state-of-the-art Hamiltonian Monte Carlo sampler, MicroCanonical Hamiltonian Monte Carlo, a recently developed sampler by \cite{robnik2022microcanonical} and~\cite{robnik2023microcanonical}, and Pathfinder, a state-of-the-art variational inference algorithm.
While \camb{} paired with \cobaya{}  required around 100,  480, and  1,600 CPU hours to analyze \planck{}, ACT and SPT-3G data respectively, \capse{} combined with the NUTS algorithm required only 0.5 CPU hours for the \planck{} analysis, one CPU hour for the ACT analysis, and 14 hours for the SPT-3G analysis. Thus our analyses were at least two orders of magnitude more efficient.
When comparing NUTS to MCHMC, we found the latter to be more efficient by a factor of 2-3, while producing virtually identical posteriors. Pathfinder was able to perform the \planck{} and ACT analysis in 1-10 s, with a very good precision. However, this approach is less robust when compared to MCMC methods as it lacks asymptotic convergence guarantees.

In this work we also tested a data reduction technique, the Chebyshev polynomial decomposition. This technique was instrumental in improving \capse{} computational performance, as it improved the \planck{} analysis efficiency by almost an order of magnitude.

While our emulators and methods exhibits high precision and computational performance, there remain opportunities for future improvements. 
First, we aim to enhance the preprocessing procedure, addressing the reduced precision of \capse{} for low-$\ell$ $EE$ 2-point function spectra by employing a proper $\tau$ rescaling.
Implementing a \textit{weighted} mean square error loss function (similar to \citet{nygaard_connect_2022}) based on the CMB power spectra variance could also lead to a more homogeneous emulation error at no additional computational cost.
We also plan to work on refining the Chebyshev polynomial decomposition approach. Although it demonstrated precision and efficiency within the \planck{} multipole range, we intend to explore ways to improve its robustness when considering higher $\ell_\mathrm{max}$ values. In particular, some preliminary tests show that we can reach the same $\ell_\mathrm{max}$ of \cosmopower{}, while maintaining the computational efficiency of our framework and even improving the precision of our emulators with respect to what we showed in this paper.

Working on the Chebyshev polynomial decomposition approach might not only be relevant for CMB analyses but also for spectroscopic galaxy clustering analyses. The estimators used to extract cosmological information from galaxy surveys~\citep{feldman_power_1994, yamamoto_measurement_2006} suffer from an imprint of the survey window geometry. In order to take into account this effect, the community has developed several methods tailored for power spectrum~\citep{Blake_2013, Beutler_2016} estimation. However, only recently this issue has been addressed for the bispectrum~\citep{pardede_bispectrum-window_2022, alkhanishvili_window_2022}. We plan to investigate whether the Chebyshev polynomial decomposition can be useful for this kind of application.

\section*{Acknowledgements}
We would like thank Boris Bolliet and Alessio Spurio Mancini for clarifications regarding \cosmopower{}.
MB is expecially grateful to Martin White for the initial suggestion on employing the Chebyshev polynomial decomposition, to Marius Millea for his advices while coding the SPT-3G likelihood in \julia{} and to Emanuele Castorina, Giulio Fabbian, SImone Ferraro and Uros Seljak for useful discussions. MB thanks University of California Berkeley for hospitality during his visit, during which this project was started.
MB acknowledges financial support from INAF MiniGrant 2022.
JRZ is supported by an STFC doctoral studentship.
F.B. acknowledges support by the Department of Energy, Contract DE-AC02-76SF00515.

\bibliographystyle{mnras}
\bibliography{biblio}

\begin{thebibliography}{}
\makeatletter
\relax
\def\mn@urlcharsother{\let\do\@makeother \do\$\do\&\do\#\do\^\do\_\do\%\do\~}
\def\mn@doi{\begingroup\mn@urlcharsother \@ifnextchar [ {\mn@doi@}
  {\mn@doi@[]}}
\def\mn@doi@[#1]#2{\def\@tempa{#1}\ifx\@tempa\@empty \href
  {http://dx.doi.org/#2} {doi:#2}\else \href {http://dx.doi.org/#2} {#1}\fi
  \endgroup}
\def\mn@eprint#1#2{\mn@eprint@#1:#2::\@nil}
\def\mn@eprint@arXiv#1{\href {http://arxiv.org/abs/#1} {{\tt arXiv:#1}}}
\def\mn@eprint@dblp#1{\href {http://dblp.uni-trier.de/rec/bibtex/#1.xml}
  {dblp:#1}}
\def\mn@eprint@#1:#2:#3:#4\@nil{\def\@tempa {#1}\def\@tempb {#2}\def\@tempc
  {#3}\ifx \@tempc \@empty \let \@tempc \@tempb \let \@tempb \@tempa \fi \ifx
  \@tempb \@empty \def\@tempb {arXiv}\fi \@ifundefined
  {mn@eprint@\@tempb}{\@tempb:\@tempc}{\expandafter \expandafter \csname
  mn@eprint@\@tempb\endcsname \expandafter{\@tempc}}}

\bibitem[\protect\citeauthoryear{Abate et~al.}{Abate
  et~al.}{2012}]{lsst_dark_energy_science_collaboration_large_2012}
Abate A.,  et~al., 2012, {Large Synoptic Survey Telescope: Dark Energy Science
  Collaboration} (\mn@eprint {arXiv} {1211.0310})

\bibitem[\protect\citeauthoryear{Abazajian et~al.,}{Abazajian
  et~al.}{2019}]{abazajian2019cmbs4}
Abazajian K.,  et~al., 2019, CMB-S4 Decadal Survey APC White Paper (\mn@eprint
  {arXiv} {1908.01062})

\bibitem[\protect\citeauthoryear{Abbott et~al.}{Abbott
  et~al.}{2018}]{abbott_dark_2018}
Abbott T. M.~C.,  et~al., 2018, \mn@doi [Phys. Rev. D]
  {10.1103/PhysRevD.98.043526}, 98, 043526

\bibitem[\protect\citeauthoryear{Ade et~al.}{Ade et~al.}{2019}]{Ade_2019}
Ade P.,  et~al., 2019, \mn@doi [JCAP] {10.1088/1475-7516/2019/02/056}, 02, 056

\bibitem[\protect\citeauthoryear{Aghanim et~al.}{Aghanim
  et~al.}{2020a}]{Planck:2019nip}
Aghanim N.,  et~al., 2020a, \mn@doi [Astron. Astrophys.]
  {10.1051/0004-6361/201936386}, 641, A5

\bibitem[\protect\citeauthoryear{Aghanim et~al.}{Aghanim
  et~al.}{2020b}]{aghanim_planck_2020}
Aghanim N.,  et~al., 2020b, \mn@doi [Astron. Astrophys.]
  {10.1051/0004-6361/201833910}, 641, A6

\bibitem[\protect\citeauthoryear{Aiola et~al.}{Aiola
  et~al.}{2020}]{aiola_atacama_2020}
Aiola S.,  et~al., 2020, \mn@doi [JCAP] {10.1088/1475-7516/2020/12/047}, 12,
  047

\bibitem[\protect\citeauthoryear{Albers, Fidler, Lesgourgues, Sch\"oneberg  \&
  Torrado}{Albers et~al.}{2019}]{albers_cosmicnet_2019}
Albers J.,  Fidler C.,  Lesgourgues J.,  Sch\"oneberg N.,   Torrado J.,  2019,
  \mn@doi [JCAP] {10.1088/1475-7516/2019/09/028}, 09, 028

\bibitem[\protect\citeauthoryear{Alkhanishvili, Porciani  \&
  Sefusatti}{Alkhanishvili et~al.}{2023}]{alkhanishvili_window_2022}
Alkhanishvili D.,  Porciani C.,   Sefusatti E.,  2023, \mn@doi [Astron.
  Astrophys.] {10.1051/0004-6361/202245156}, 669, L2

\bibitem[\protect\citeauthoryear{Alvarez, Ferraro, Hill, Hlo\v{z}ek  \&
  Ikape}{Alvarez et~al.}{2021}]{Alvarez:2020gvl}
Alvarez M.~A.,  Ferraro S.,  Hill J.~C.,  Hlo\v{z}ek R.,   Ikape M.,  2021,
  \mn@doi [Phys. Rev. D] {10.1103/PhysRevD.103.063518}, 103, 063518

\bibitem[\protect\citeauthoryear{Angulo, Zennaro, Contreras, Aric\`o,
  Pellejero-Iba\~nez  \& St\"ucker}{Angulo et~al.}{2021}]{angulo_bacco_2021}
Angulo R.~E.,  Zennaro M.,  Contreras S.,  Aric\`o G.,  Pellejero-Iba\~nez M.,
   St\"ucker J.,  2021, \mn@doi [Mon. Not. Roy. Astron. Soc.]
  {10.1093/mnras/stab2018}, 507, 5869

\bibitem[\protect\citeauthoryear{Aric\`o, Angulo  \& Zennaro}{Aric\`o
  et~al.}{2021}]{arico_accelerating_2021}
Aric\`o G.,  Angulo R.~E.,   Zennaro M.,  2021, \mn@doi []
  {10.12688/openreseurope.14310.2}

\bibitem[\protect\citeauthoryear{Auld, Bridges, Hobson  \& Gull}{Auld
  et~al.}{2007}]{auld_fast_2007}
Auld T.,  Bridges M.,  Hobson M.~P.,   Gull S.~F.,  2007, \mn@doi [Mon. Not.
  Roy. Astron. Soc.] {10.1111/j.1745-3933.2006.00276.x}, 376, L11

\bibitem[\protect\citeauthoryear{Balkenhol et~al.}{Balkenhol
  et~al.}{2023}]{balkenhol23}
Balkenhol L.,  et~al., 2023, \mn@doi [Phys. Rev. D]
  {10.1103/PhysRevD.108.023510}, 108, 023510

\bibitem[\protect\citeauthoryear{Bayer, Seljak  \& Modi}{Bayer
  et~al.}{2023}]{Bayer:2023rmj}
Bayer A.~E.,  Seljak U.,   Modi C.,  2023, in {40th International Conference on
  Machine Learning}.  (\mn@eprint {arXiv} {2307.09504})

\bibitem[\protect\citeauthoryear{Betancourt}{Betancourt}{2018}]{betancourt2018conceptual}
Betancourt M.,  2018, A Conceptual Introduction to Hamiltonian Monte Carlo
  (\mn@eprint {arXiv} {1701.02434})

\bibitem[\protect\citeauthoryear{Beutler et~al.}{Beutler
  et~al.}{2017}]{Beutler_2016}
Beutler F.,  et~al., 2017, \mn@doi [Mon. Not. Roy. Astron. Soc.]
  {10.1093/mnras/stw3298}, 466, 2242

\bibitem[\protect\citeauthoryear{Bezanson, Edelman, Karpinski  \&
  Shah}{Bezanson et~al.}{2015}]{bezanson_julia_2015}
Bezanson J.,  Edelman A.,  Karpinski S.,   Shah V.~B.,  2015, Julia: A Fresh
  Approach to Numerical Computing (\mn@eprint {arXiv} {1411.1607})

\bibitem[\protect\citeauthoryear{Blake et~al.}{Blake et~al.}{2013}]{Blake_2013}
Blake C.,  et~al., 2013, \mn@doi [Mon. Not. Roy. Astron. Soc.]
  {10.1093/mnras/stt1791}, 436, 3089

\bibitem[\protect\citeauthoryear{Blas, Lesgourgues  \& Tram}{Blas
  et~al.}{2011}]{blas_cosmic_2011}
Blas D.,  Lesgourgues J.,   Tram T.,  2011, \mn@doi [JCAP]
  {10.1088/1475-7516/2011/07/034}, 07, 034

\bibitem[\protect\citeauthoryear{Bolliet, Spurio~Mancini, Hill, Madhavacheril,
  Jense, Calabrese  \& Dunkley}{Bolliet
  et~al.}{2023}]{bolliet_high-accuracy_2023}
Bolliet B.,  Spurio~Mancini A.,  Hill J.~C.,  Madhavacheril M.,  Jense H.~T.,
  Calabrese E.,   Dunkley J.,  2023, {High-accuracy emulators for observables
  in $\Lambda$CDM, $N_\mathrm{eff}$, $\Sigma m_\nu$, and $w$ cosmologies}
  (\mn@eprint {arXiv} {2303.01591})

\bibitem[\protect\citeauthoryear{Bonici, Biggio, Carbone  \& Guzzo}{Bonici
  et~al.}{2022}]{bonici_fast_2022}
Bonici M.,  Biggio L.,  Carbone C.,   Guzzo L.,  2022, {Fast emulation of
  two-point angular statistics for photometric galaxy surveys} (\mn@eprint
  {arXiv} {2206.14208})

\bibitem[\protect\citeauthoryear{Brooks, Gelman, Jones  \& Meng}{Brooks
  et~al.}{2011}]{brooks2011handbook}
Brooks S.,  Gelman A.,  Jones G.,   Meng X.-L.,  2011, Handbook of Markov Chain
  Monte Carlo.
CRC press

\bibitem[\protect\citeauthoryear{Campagne et~al.,}{Campagne
  et~al.}{2023}]{Campagne:2023ter}
Campagne J.-E.,  et~al., 2023, \mn@doi [Open J. Astrophys.]
  {10.21105/astro.2302.05163}, 6, 1

\bibitem[\protect\citeauthoryear{Crill et~al.,}{Crill
  et~al.}{2020}]{crill_spherex_2020}
Crill B.~P.,  et~al., 2020, in Space {Telescopes} and {Instrumentation} 2020:
  {Optical}, {Infrared}, and {Millimeter} {Wave}. SPIE, pp 61--77,
  \mn@doi{10.1117/12.2567224}, \url
  {https://www.spiedigitallibrary.org/conference-proceedings-of-spie/11443/114430I/SPHEREx-NASAs-near-infrared-spectrophotometric-all-sky-survey/10.1117/12.2567224.full}

\bibitem[\protect\citeauthoryear{Dawson et~al.}{Dawson
  et~al.}{2013}]{dawson_baryon_2013}
Dawson K.~S.,  et~al., 2013, \mn@doi [Astron. J.] {10.1088/0004-6256/145/1/10},
  145, 10

\bibitem[\protect\citeauthoryear{DeRose et~al.,}{DeRose
  et~al.}{2019}]{DeRose_2019}
DeRose J.,  et~al., 2019, \mn@doi [Astrophys. J.] {10.3847/1538-4357/ab1085},
  875, 69

\bibitem[\protect\citeauthoryear{Donald-McCann, Beutler, Koyama  \&
  Karamanis}{Donald-McCann et~al.}{2022a}]{donald-mccann_textttmatryoshka_2021}
Donald-McCann J.,  Beutler F.,  Koyama K.,   Karamanis M.,  2022a, \mn@doi
  [Mon. Not. Roy. Astron. Soc.] {10.1093/mnras/stac239}, 511, 3768

\bibitem[\protect\citeauthoryear{Donald-McCann, Koyama  \&
  Beutler}{Donald-McCann et~al.}{2022b}]{donald-mccann_textttmatryoshka_2022}
Donald-McCann J.,  Koyama K.,   Beutler F.,  2022b, \mn@doi [Mon. Not. Roy.
  Astron. Soc.] {10.1093/mnras/stac3326}, 518, 3106

\bibitem[\protect\citeauthoryear{Dutcher et~al.}{Dutcher
  et~al.}{2021}]{dutcher21}
Dutcher D.,  et~al., 2021, \mn@doi [Phys. Rev. D]
  {10.1103/PhysRevD.104.022003}, 104, 022003

\bibitem[\protect\citeauthoryear{Eggemeier, Camacho-Quevedo, Pezzotta, Crocce,
  Scoccimarro  \& S\'anchez}{Eggemeier et~al.}{2022}]{eggemeier_comet_2022}
Eggemeier A.,  Camacho-Quevedo B.,  Pezzotta A.,  Crocce M.,  Scoccimarro R.,
  S\'anchez A.~G.,  2022, \mn@doi [Mon. Not. Roy. Astron. Soc.]
  {10.1093/mnras/stac3667}, 519, 2962

\bibitem[\protect\citeauthoryear{Feldman, Kaiser  \& Peacock}{Feldman
  et~al.}{1994}]{feldman_power_1994}
Feldman H.~A.,  Kaiser N.,   Peacock J.~A.,  1994, \mn@doi [Astrophys. J.]
  {10.1086/174036}, 426, 23

\bibitem[\protect\citeauthoryear{Fendt \& Wandelt}{Fendt \&
  Wandelt}{2006}]{fendt_pico_2007}
Fendt W.~A.,  Wandelt B.~D.,  2006, \mn@doi [Astrophys. J.] {10.1086/508342},
  654, 2

\bibitem[\protect\citeauthoryear{Gammal, Sch\"oneberg, Torrado  \&
  Fidler}{Gammal et~al.}{2022}]{Gammal:2022eob}
Gammal J.~E.,  Sch\"oneberg N.,  Torrado J.,   Fidler C.,  2022, {Fast and
  robust Bayesian Inference using Gaussian Processes with GPry} (\mn@eprint
  {arXiv} {2211.02045})

\bibitem[\protect\citeauthoryear{Ge, Xu  \& Ghahramani}{Ge
  et~al.}{2018}]{ge2018t}
Ge H.,  Xu K.,   Ghahramani Z.,  2018, in International Conference on
  Artificial Intelligence and Statistics, {AISTATS} 2018, 9-11 April 2018,
  Playa Blanca, Lanzarote, Canary Islands, Spain. pp 1682--1690, \url
  {http://proceedings.mlr.press/v84/ge18b.html}

\bibitem[\protect\citeauthoryear{G\"unther}{G\"unther}{2023}]{Gunther:2023xhh}
G\"unther S.,  2023, {Uncertainty-aware and Data-efficient Cosmological
  Emulation using Gaussian Processes and PCA} (\mn@eprint {arXiv} {2307.01138})

\bibitem[\protect\citeauthoryear{G\"unther, Lesgourgues, Samaras, Sch\"oneberg,
  Stadtmann, Fidler  \& Torrado}{G\"unther
  et~al.}{2022}]{gunther_cosmicnet_2022}
G\"unther S.,  Lesgourgues J.,  Samaras G.,  Sch\"oneberg N.,  Stadtmann F.,
  Fidler C.,   Torrado J.,  2022, \mn@doi [JCAP]
  {10.1088/1475-7516/2022/11/035}, 11, 035

\bibitem[\protect\citeauthoryear{Heavens, Jimenez  \& Lahav}{Heavens
  et~al.}{2000}]{Heavens:1999am}
Heavens A.,  Jimenez R.,   Lahav O.,  2000, \mn@doi [Mon. Not. Roy. Astron.
  Soc.] {10.1046/j.1365-8711.2000.03692.x}, 317, 965

\bibitem[\protect\citeauthoryear{Heavens, Sellentin, de Mijolla  \&
  Vianello}{Heavens et~al.}{2017}]{Heavens:2017efz}
Heavens A.,  Sellentin E.,  de Mijolla D.,   Vianello A.,  2017, \mn@doi [Mon.
  Not. Roy. Astron. Soc.] {10.1093/mnras/stx2326}, 472, 4244

\bibitem[\protect\citeauthoryear{Hoffman \& Gelman}{Hoffman \&
  Gelman}{2011}]{hoffman2011nouturn}
Hoffman M.~D.,  Gelman A.,  2011, The No-U-Turn Sampler: Adaptively Setting
  Path Lengths in Hamiltonian Monte Carlo (\mn@eprint {arXiv} {1111.4246})

\bibitem[\protect\citeauthoryear{Jimenez, Verde, Peiris  \& Kosowsky}{Jimenez
  et~al.}{2004}]{jimenez_fast_2004}
Jimenez R.,  Verde L.,  Peiris H.,   Kosowsky A.,  2004, \mn@doi [Phys. Rev. D]
  {10.1103/PhysRevD.70.023005}, 70, 023005

\bibitem[\protect\citeauthoryear{Kamionkowski, Kosowsky  \&
  Stebbins}{Kamionkowski et~al.}{1997}]{Kamionkowski:1996ks}
Kamionkowski M.,  Kosowsky A.,   Stebbins A.,  1997, \mn@doi [Phys. Rev. D]
  {10.1103/PhysRevD.55.7368}, 55, 7368

\bibitem[\protect\citeauthoryear{Kingma \& Ba}{Kingma \&
  Ba}{2017}]{kingma_adam_2017}
Kingma D.~P.,  Ba J.,  2017, Adam: A Method for Stochastic Optimization
  (\mn@eprint {arXiv} {1412.6980})

\bibitem[\protect\citeauthoryear{Knabenhans et~al.}{Knabenhans
  et~al.}{2019}]{euclid_collaboration_euclid_2019}
Knabenhans M.,  et~al., 2019, \mn@doi [Mon. Not. Roy. Astron. Soc.]
  {10.1093/mnras/stz197}, 484, 5509

\bibitem[\protect\citeauthoryear{Knabenhans et~al.}{Knabenhans
  et~al.}{2021}]{euclidcollaboration_euclid_2021}
Knabenhans M.,  et~al., 2021, \mn@doi [Mon. Not. Roy. Astron. Soc.]
  {10.1093/mnras/stab1366}, 505, 2840

\bibitem[\protect\citeauthoryear{Knox}{Knox}{1995}]{Knox:1995dq}
Knox L.,  1995, \mn@doi [Phys. Rev. D] {10.1103/PhysRevD.52.4307}, 52, 4307

\bibitem[\protect\citeauthoryear{Kucukelbir, Tran, Ranganath, Gelman  \&
  Blei}{Kucukelbir et~al.}{2016}]{kucukelbir2016automatic}
Kucukelbir A.,  Tran D.,  Ranganath R.,  Gelman A.,   Blei D.~M.,  2016,
  Automatic Differentiation Variational Inference (\mn@eprint {arXiv}
  {1603.00788})

\bibitem[\protect\citeauthoryear{Laureijs et~al.}{Laureijs
  et~al.}{2011}]{laureijs_euclid_2011}
Laureijs R.,  et~al., 2011, {Euclid Definition Study Report} (\mn@eprint
  {arXiv} {1110.3193})

\bibitem[\protect\citeauthoryear{Lawrence et~al.,}{Lawrence
  et~al.}{2017}]{Lawrence_2017}
Lawrence E.,  et~al., 2017, \mn@doi [Astrophys. J.] {10.3847/1538-4357/aa86a9},
  847, 50

\bibitem[\protect\citeauthoryear{Levi et~al.}{Levi
  et~al.}{2013}]{levi_desi_2013}
Levi M.,  et~al., 2013, {The DESI Experiment, a whitepaper for Snowmass 2013}
  (\mn@eprint {arXiv} {1308.0847})

\bibitem[\protect\citeauthoryear{Lewis, Challinor  \& Lasenby}{Lewis
  et~al.}{2000}]{lewis_efficient_2000}
Lewis A.,  Challinor A.,   Lasenby A.,  2000, \mn@doi [Astrophys. J.]
  {10.1086/309179}, 538, 473

\bibitem[\protect\citeauthoryear{Manrique-Yus \& Sellentin}{Manrique-Yus \&
  Sellentin}{2020}]{manrique-yus_euclid-era_2019}
Manrique-Yus A.,  Sellentin E.,  2020, \mn@doi [Mon. Not. Roy. Astron. Soc.]
  {10.1093/mnras/stz3059}, 491, 2655

\bibitem[\protect\citeauthoryear{McKay, Beckman  \& Conover}{McKay
  et~al.}{1979}]{McKay1979}
McKay M.~D.,  Beckman R.~J.,   Conover W.~J.,  1979, Technometrics, 21, 239

\bibitem[\protect\citeauthoryear{Mead, Brieden, Tr\"oster  \& Heymans}{Mead
  et~al.}{2020}]{mead_hmcode-2020_2021}
Mead A.,  Brieden S.,  Tr\"oster T.,   Heymans C.,  2020, \mn@doi [Mon. Not.
  Roy. Astron. Soc.] {10.1093/mnras/stab082}

\bibitem[\protect\citeauthoryear{Mootoovaloo, Heavens, Jaffe  \&
  Leclercq}{Mootoovaloo et~al.}{2020}]{mootoovaloo_parameter_2020}
Mootoovaloo A.,  Heavens A.~F.,  Jaffe A.~H.,   Leclercq F.,  2020, \mn@doi
  [Mon. Not. Roy. Astron. Soc.] {10.1093/mnras/staa2102}, 497, 2213

\bibitem[\protect\citeauthoryear{Mootoovaloo, Jaffe, Heavens  \&
  Leclercq}{Mootoovaloo et~al.}{2022}]{Mootoovaloo:2021rot}
Mootoovaloo A.,  Jaffe A.~H.,  Heavens A.~F.,   Leclercq F.,  2022, \mn@doi
  [Astron. Comput.] {10.1016/j.ascom.2021.100508}, 38, 100508

\bibitem[\protect\citeauthoryear{Nygaard, Holm, Hannestad  \& Tram}{Nygaard
  et~al.}{2023}]{nygaard_connect_2022}
Nygaard A.,  Holm E.~B.,  Hannestad S.,   Tram T.,  2023, \mn@doi [JCAP]
  {10.1088/1475-7516/2023/05/025}, 05, 025

\bibitem[\protect\citeauthoryear{Pardede, Rizzo, Biagetti, Castorina, Sefusatti
   \& Monaco}{Pardede et~al.}{2022}]{pardede_bispectrum-window_2022}
Pardede K.,  Rizzo F.,  Biagetti M.,  Castorina E.,  Sefusatti E.,   Monaco P.,
   2022, \mn@doi [JCAP] {10.1088/1475-7516/2022/10/066}, 10, 066

\bibitem[\protect\citeauthoryear{Piras \& Spurio~Mancini}{Piras \&
  Spurio~Mancini}{2023}]{Piras_2023}
Piras D.,  Spurio~Mancini A.,  2023, \mn@doi [Open J. Astrophys.]
  {10.21105/astro.2305.06347}

\bibitem[\protect\citeauthoryear{Prince \& Dunkley}{Prince \&
  Dunkley}{2019}]{prince_data_2019}
Prince H.,  Dunkley J.,  2019, \mn@doi [Phys. Rev. D]
  {10.1103/PhysRevD.100.083502}, 100, 083502

\bibitem[\protect\citeauthoryear{Ranganath, Gerrish  \& Blei}{Ranganath
  et~al.}{2013}]{ranganath2013black}
Ranganath R.,  Gerrish S.,   Blei D.~M.,  2013, Black Box Variational Inference
  (\mn@eprint {arXiv} {1401.0118})

\bibitem[\protect\citeauthoryear{Robnik \& Seljak}{Robnik \&
  Seljak}{2023}]{robnik2023microcanonical}
Robnik J.,  Seljak U.,  2023, Microcanonical Langevin Monte Carlo (\mn@eprint
  {arXiv} {2303.18221})

\bibitem[\protect\citeauthoryear{Robnik, Luca, Silverstein  \& Seljak}{Robnik
  et~al.}{2022}]{robnik2022microcanonical}
Robnik J.,  Luca G. B.~D.,  Silverstein E.,   Seljak U.,  2022, Microcanonical
  Hamiltonian Monte Carlo (\mn@eprint {arXiv} {2212.08549})

\bibitem[\protect\citeauthoryear{Sobrin et~al.}{Sobrin
  et~al.}{2022}]{Sobrin_2022}
Sobrin J.~A.,  et~al., 2022, \mn@doi [Astrophys. J. Supp.]
  {10.3847/1538-4365/ac374f}, 258, 42

\bibitem[\protect\citeauthoryear{Spergel et~al.}{Spergel
  et~al.}{2015}]{spergel_wide-field_2015}
Spergel D.,  et~al., 2015, {Wide-Field InfrarRed Survey Telescope-Astrophysics
  Focused Telescope Assets WFIRST-AFTA 2015 Report} (\mn@eprint {arXiv}
  {1503.03757})

\bibitem[\protect\citeauthoryear{Spurio~Mancini, Piras, Alsing, Joachimi  \&
  Hobson}{Spurio~Mancini et~al.}{2022}]{mancini_itcosmopower_2021}
Spurio~Mancini A.,  Piras D.,  Alsing J.,  Joachimi B.,   Hobson M.~P.,  2022,
  \mn@doi [Mon. Not. Roy. Astron. Soc.] {10.1093/mnras/stac064}, 511, 1771

\bibitem[\protect\citeauthoryear{To, Rozo, Krause, Wu, Wechsler  \& Salcedo}{To
  et~al.}{2023}]{To:2022ubu}
To C.-H.,  Rozo E.,  Krause E.,  Wu H.-Y.,  Wechsler R.~H.,   Salcedo A.~N.,
  2023, \mn@doi [JCAP] {10.1088/1475-7516/2023/01/016}, 01, 016

\bibitem[\protect\citeauthoryear{Trefethen}{Trefethen}{2012}]{atap}
Trefethen L.~N.,  2012, Approximation Theory and Approximation Practice..
SIAM

\bibitem[\protect\citeauthoryear{Vehtari, Simpson, Gelman, Yao  \&
  Gabry}{Vehtari et~al.}{2022}]{vehtari2022pareto}
Vehtari A.,  Simpson D.,  Gelman A.,  Yao Y.,   Gabry J.,  2022, Pareto
  Smoothed Importance Sampling (\mn@eprint {arXiv} {1507.02646})

\bibitem[\protect\citeauthoryear{Wang}{Wang}{2023}]{wang2023analysis}
Wang H.,  2023, Analysis of error localization of Chebyshev spectral
  approximations (\mn@eprint {arXiv} {2106.03456})

\bibitem[\protect\citeauthoryear{Yamamoto, Nakamichi, Kamino, Bassett  \&
  Nishioka}{Yamamoto et~al.}{2006}]{yamamoto_measurement_2006}
Yamamoto K.,  Nakamichi M.,  Kamino A.,  Bassett B.~A.,   Nishioka H.,  2006,
  \mn@doi [Publ. Astron. Soc. Jap.] {10.1093/pasj/58.1.93}, 58, 93

\bibitem[\protect\citeauthoryear{Zhang, Carpenter, Gelman  \& Vehtari}{Zhang
  et~al.}{2022}]{zhang2022pathfinder}
Zhang L.,  Carpenter B.,  Gelman A.,   Vehtari A.,  2022, Pathfinder: Parallel
  quasi-Newton variational inference (\mn@eprint {arXiv} {2108.03782})

\bibitem[\protect\citeauthoryear{de Jong et~al.}{de~Jong
  et~al.}{}]{de_jong_kilo-degree_2013}
de Jong J. T.~A.,  et~al.,, {The Kilo-Degree Survey}

\makeatother
\end{thebibliography}


\appendix
\section{Affine transformation of Multivariate distribution}
\label{app:reparametrization}
Let us consider a random variable $X$, distributed according to:
\begin{equation}
    X \sim \mathcal{N}(\mu_X,\Sigma_X).
\end{equation}
If we define a new variable:
\begin{equation}
    Y = LX+u,
\end{equation}
then $Y$ follows:
\begin{equation}
    Y\sim \mathcal{N}(\mu_Y,\Sigma_Y)=\mathcal{N}(u+L\mu_X,L\Sigma_X L^T).
    \label{eq:reparametrization}
\end{equation}
This can be proved as follows:
\begin{equation}
    \mu_Y=\mathbb{E}[Y]=\mathbb{E}[L X+u]=\mathbb{E}[L X]+u=L \mu_X+u
\end{equation}
\begin{equation}
    \begin{aligned} \Sigma_Y & =\mathbb{E}\left[\left(Y-\mu_Y\right)\left(Y-\mu_Y\right)^{\top}\right] \\ & =\mathbb{E}\left[\left(L X+u-L \mu_X-u\right)\left(L X+u-L \mu_X-u\right)^{\top}\right] \\ & =\mathbb{E}\left[\left(L\left(X-\mu_X\right)\right)\left(L\left(X-\mu_X\right)\right)^{\top}\right] \\ & =\mathbb{E}\left[L\left(X-\mu_X\right)\left(X-\mu_X\right)^{\top} L^{\top}\right] \\ & =L \mathbb{E}\left[\left(X-\mu_X\right)\left(X-\mu_X\right)^{\top}\right] L^{\top} \\ & =L \Sigma_X L^{\top}.\end{aligned}
\end{equation}

In this work we consider likelihood of the form:
\begin{equation}
    d\sim\mathcal{N}\left(t(\theta), \Sigma\right)
\end{equation}
where $d$ represents either \planck{} or ACT data and $\Sigma$ their covariance.
Given the \capse{} emulator performance, computing the inverse of the covariance matrix gives a not negligible computational overhead. Eq.~\ref{eq:reparametrization} is interesting for our discussion as it gives us a way to avoid computing the inverse of the covariance matrix at each step of the MonteCarlo. In particular, if we define:
\begin{equation}
    \Gamma = \mathrm{sqrt}(\Sigma),
\end{equation}
\begin{equation}
    \mathcal{D} = \mathrm{inv}(\Gamma)d,
\end{equation}
then:
\begin{equation}
    \mathcal{D} \sim \mathcal{N}(\mathrm{inv}(\Gamma)t(\theta), \mathds{I}).
\end{equation}
We use this parametrization in our likelihoods.

\newpage
\section{Simons Observatory residuals distribution}
\begin{figure*}[ht]%
\centering%
\includegraphics[width=.85\textwidth]{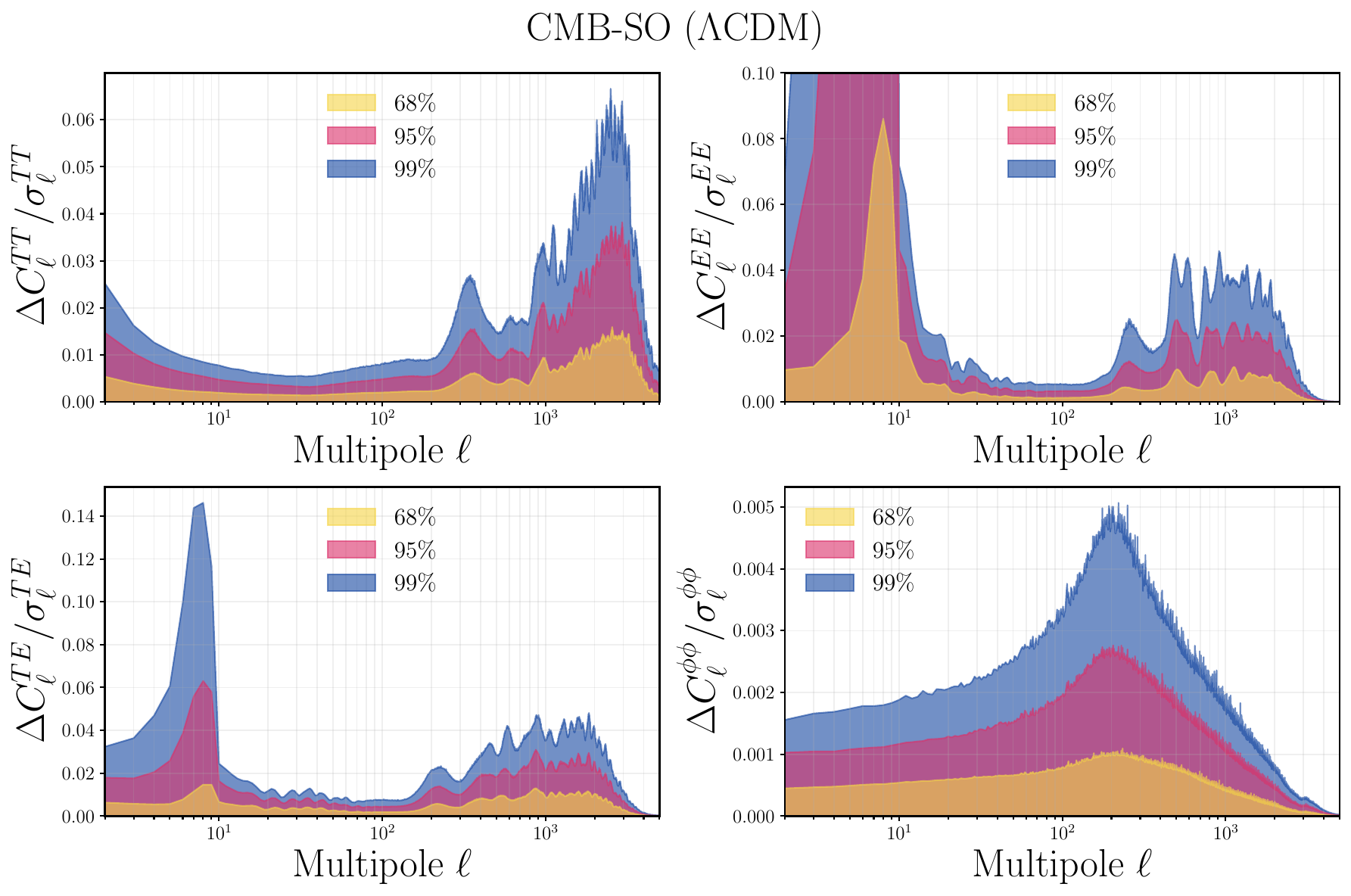}%
\caption{Distribution of the relative difference between \texttt{Capse.jl} and the high-accuracy predictions from \texttt{CAMB}  across the four 2-point statistics emulated in this work, normalized to the forecasted SO statistical uncertainties ($f_\mathrm{sky}=0.4$). The differences are in units of cosmic variance below $\ell < 30$. The shaded regions denote the 68\%, 95\%, and 99\% percentiles of the distributions, respectively.}%
\label{fig:SO_errors}
\end{figure*}

\newpage
\section{\planck{} MOPED likelihood contours}
\begin{figure*}[ht]%
\centering%
\includegraphics[width=\textwidth]{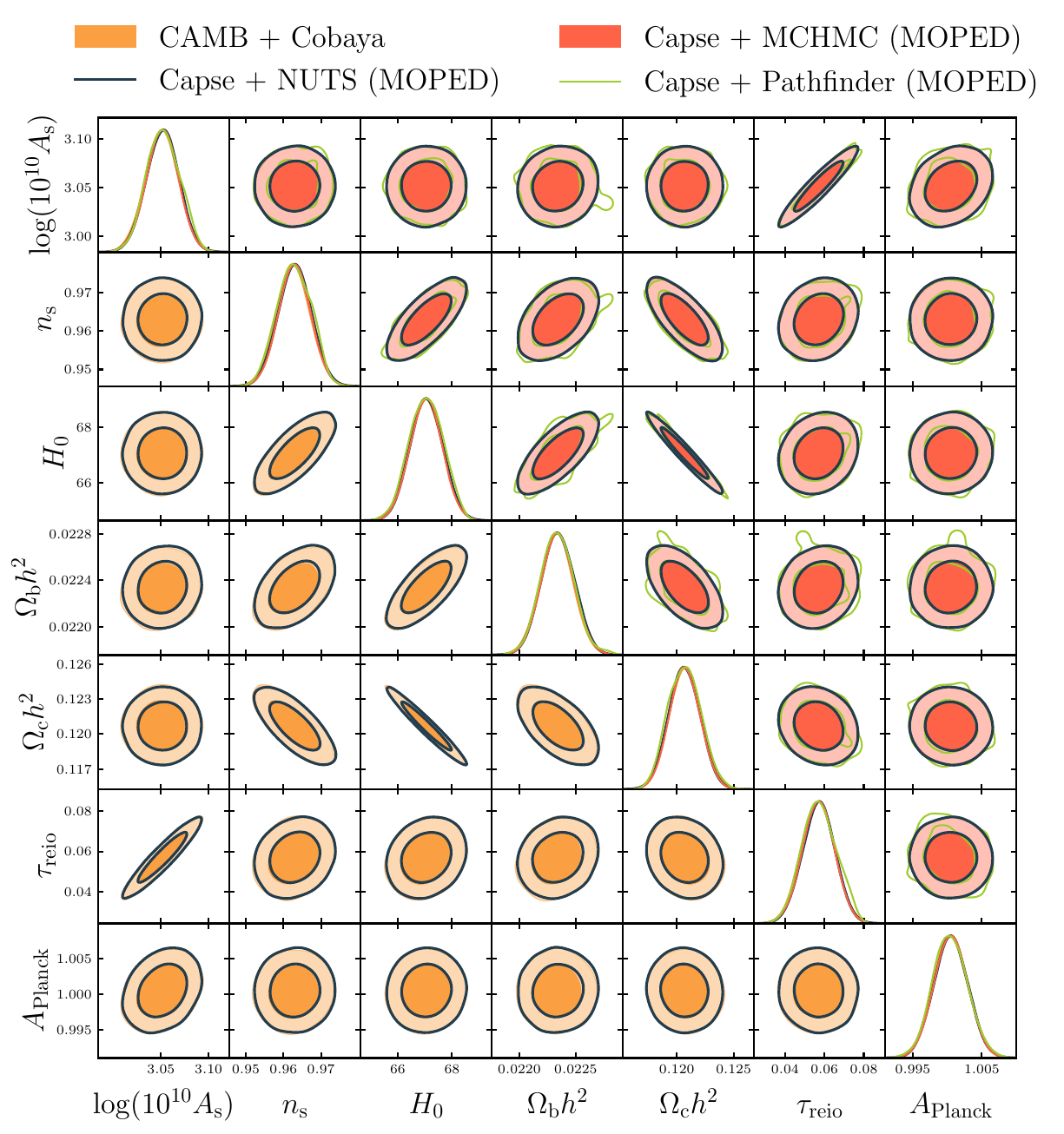}%
\caption{Contours obtained analyzing the \planck{} premarginalized dataset. As in Fig.~\ref{fig:contour_planck},  we compare the \camb{} chains with the \capse{} ones, the main difference being that we used the MOPED likelihood for the emulator-based runs. Also in this case, the comparison of the contours is very good, being the difference on marginalized parameters smaller than $0.1\,\sigma$.}%
\label{fig:planck_MOPED}
\end{figure*}

\end{document}